\documentclass[twocolumn]{aastex63}
\usepackage{gensymb}    
\usepackage{bm}

\newcommand{\bdm}{\begin{displaymath}}
\newcommand{\edm}{\end{displaymath}}
\newcommand{\beq}{\begin{equation}}
\newcommand{\eeq}{\end{equation}}

\renewcommand{\vec}[1]{\bm{#1}}
\renewcommand{\tensor}[1]{\mathbf{#1}}
\newcommand{\nablab}{\bm{\nabla}}
\newcommand{\cdotb}{\bm{\cdot}}
\newcommand{\cross}{\bm{\times}}

\definecolor{remred}{rgb}{0.804, 0.31, 0.224}
\definecolor{addgreen}{rgb}{0.27, 0.545, 0}
\iftrue
    \newcommand{\cadded}[1]{#1}
    \newcommand{\cremoved}[1]{}
    \newcommand{\creplaced}[2]{#2}
\fi

\received{August 4$^{\rm th}$, 2020}
\revised{October 1$^{\rm st}$, 2020}
\accepted{October 2$^{\rm nd}$, 2020}

\submitjournal{ApJ}

\shorttitle{Semi-Detached Binaries in MHD}
\shortauthors{P. Pjanka and J. M. Stone}

\graphicspath{{./}{figures/}}

\begin{document}

\title{Stratified Global MHD Models of Accretion Disks\\ in Semi-Detached Binaries}

\correspondingauthor{Patryk Pjanka}
\email{patryk.pjanka@su.se}

\author[0000-0003-3564-9689]{Patryk Pjanka}
\affiliation{Department of Astrophysical Sciences, Princeton University, 4 Ivy Lane, Princeton, NJ 08544, USA}
\affiliation{Nordita, KTH Royal Institute of Technology and Stockholm University, Roslagstullsbacken 23, SE-10691 Stockholm, Sweden}

\author{James M. Stone}
\affiliation{Institute for Advanced Study, 1 Einstein Drive, Princeton, NJ 08540, USA}
\affiliation{Department of Astrophysical Sciences, Princeton University, 4 Ivy Lane, Princeton, NJ 08544, USA}


\begin{abstract}
We present results of the first global magnetohydrodynamic (MHD) simulations of accretion disks fed by Roche lobe overflow, including vertical stratification, in order to investigate the roles of spiral shocks, magnetorotational instability (MRI), and the accretion stream on disk structure and evolution. Our models include a simple treatment of gas thermodynamics, with orbital Mach numbers at the inner edge of the disk $M_{\rm in}$ of 5 and 10. We find mass accretion rates to vary considerably on all time scales, with only the Mach~5 model reaching a clear quasi-stationary state. For Mach~10, the model undergoes an outside-in magnetically-driven accretion event occurring on a time scale of $\sim10$ orbital periods of the binary. Both models exhibit spiral shocks inclined with respect to the binary plane, with their position and inclination changing rapidly. However, the time-averaged location of these shocks in the equatorial plane is well-fit by simple linear models. MRI turbulence in the disk generates toroidal magnetic field patterns (butterfly diagrams) that are in some cases irregular, perhaps due to interaction with spiral structure. While many of our results are in good agreement with local studies, we find some features (most notably those related to spiral shocks) can only be captured in global models such as studied here. Thus, while global studies remain computationally expensive -- even as idealized models -- they are essential (along with more sophisticated treatment of radiation transport and disk thermodynamics) for furthering our understanding of accretion in binary systems.
\end{abstract}




\section{Introduction}\label{sect:intro}

Semi-detached binaries are some of the most interesting sources for studies of disk accretion, due to their well-defined Roche-lobe overflow mass supply, variety of their observed behavior, as well as favorable distances and numbers allowing for a wealth of observational data to be accessible for a large number of sources. These properties make them natural targets for models of disk accretion, both in a local and global sense.

Of these systems, Cataclysmic Variables (CVs) are perhaps most accessible to numerical modeling, and a large body of computational studies of their accretion flows has been accrued over the last decades. CVs are close interacting binary systems composed of a Roche-lobe-filling (usually main sequence) star and a (higher-mass) white dwarf (WD). The binary separation is generally at a few solar radii and binary periods are of the order of a few hours \citep{warner_1995}. Mass exchange in CVs leads to formation of an accretion disk around the WD resulting in a range of observational phenomena of interest in context of accretion physics. The best known of these are the recurring dwarf nova (DN) outbursts, occasionally increasing brightness of some CVs by up to $\sim8$~mag for $2-20$~days \citep{2001Lasota}. CV accretion disks may also exhibit quasi-periodic oscillations \citep[QPOs,][]{2003Warner}, as well as flickering visible in their rapid photometry \citep[e.g.,][]{1992Bruch, 2001Sokoloski, 2004Woudt}. The geometric structure of the accretion disk can be studied via imaging techniques such as eclipse mapping and Doppler tomography \citep[e.g.,][]{1997Steeghs, 2005Baptista, 2006Klingler, 2008Khruzina, 2020RuizCarmona}. 

\subsection{Local models}

Understanding the physics of angular momentum transport and acceretion in disks has often made use of the local shearing box approximation \citep{1965Goldreich, 1995Hawley}.  Many of the results from shearing box simulations are of particular interest in context of accretion in CVs, especially regarding the DN mechanism and its relation to the magnetorotational instability \citep[MRI;][]{1992HawleyBalbus}, which is now understood to be the mechanism responsible for angular momentum transport and accretion in fully-ionized plasmas. DN outbursts are thought to result from thermal instability within the disk \citep[the disk instability model, or DIM; e.g.,][]{1971Smak, 1974Osaki, 1981Meyer, 1983Faulkner, 2001Lasota}, causing hysteresis between cold and hot stable accretion branches with different hydrogen ionization levels. Assuming \cite{ShakuraSunyaev} model, DIM requires $\alpha\sim0.1$ and $\alpha\sim0.01$ in outburst and quiescence, respectively \citep[][and others]{1983Mineshige, 1984Meyer, 1984Smak}. However, connecting these values to the behavior of MRI turbulence has proven challenging. \cite{1998Gammie} suggested that lower ionization in the cold branch may cause resistivity to reduce angular momentum transport through MRI. \cite{2012Latter} were able to reproduce the two thermal states corresponding to these changes in their unstratified shearing box simulations. Adding more sophisticated ionization and opacity prescriptions improved matters further. \cite{2014Hirose, 2016Coleman, 2018Scepi_conv, 2018Coleman} were able to reproduce the characteristics of DN outbursts in their models, seeing their equilibrium states align with the expected hysteresis S-curve. They found convection to be an important mechanism increasing MRI-related accretion levels to $\alpha\sim0.1-0.2$ required to match observations \citep[see also][]{2015Hirose}.

\subsection[The challenges of global modeling in semi-detached binaries]{The challenges of global modeling\\ in semi-detached binaries}\label{sect:intro:limitations}

At first glance, CV systems appear approachable for global numerical modeling. One of the difficulties of global accretion models comes from the dynamical range of a system. In grid models, the integration time~step needs to be small enough to resolve Keplerian motion of the gas at the inner edge ($r_{\rm in}$) of the grid. At the same time, the total simulation time must be large enough to contain a number of Keplerian orbits at the outer edge of the disk ($r_{\rm out}$). The larger the ratio of the two, the more expensive a model is. CVs exhibit fairly accessible dynamic ranges relative to other disk-accretion environments, with $r_{\rm out}/r_{\rm in}\sim50-100$. 

However, a faithful representation would also need to resolve typical length- and time-scales of all relevant processes. In case of CV accretion disks, the main length-scales of concern are the disk thermal scale height and the most unstable MRI wavelength. The sonic Mach number at the inner disk edge $M_{\rm in}$ ranges in CVs from $\sim 50-200$ in outburst to $\sim 200-600$ in quiescence (see discussion in \citealt{WendyThesis}). This corresponds to the disk aspect ratio (of thermal scale height to radius) $H/R\sim1/M_{\rm in}$. Meanwhile, the mid-plane most unstable MRI wavelength is self-consistently set by the MRI dynamo (\citealt{1995Brandenburg}; \creplaced{\citealt{1995Hawley}}{\citealt{1996Hawley}}) and it is typically found in stratified shearing box studies to be $\lambda_{\rm MRI}\sim0.1-1H$ \citep[e.g.,][]{2013Mamatsashvili, 2018Scepi, 2019Suzuki}. This further increases requirements on grid resolution -- especially given specific needs of resolving MRI turbulence \citep{2011Hawley, 2012Sorathia, 2013Hawley} -- to \textit{thousands} of cells per radian. As a result, realistic values of temperature and related disk aspect ratios remain extremely challenging for numerical studies, especially in the global and/or stratified context. It is thus common to perform simulations at lower Mach numbers \citep[usually up to $\sim20$, e.g.,][]{2008Kley, WendyThesis, 2018Arzamasskiy} and apply the understanding built with these models to gain insights about more realistic regimes of temperatures.

To simulate CV disks completely rigorously, one would also need to account for the magnetic diffusion and molecular viscosity time-scales. The latter easily reaches $10^7\textrm{ yr}\simeq 10^{10}P_{\rm orb}$ (orbital periods of the binary, assuming kinetic viscosity $\nu\sim10^5\textrm{ cm}^2\textrm{s}^{-1}$ and disk radius as the typical length scale $\sim 10^{10}$~cm; \citealt{2003Balbus, 2006Johnson}). Magnetic diffusion time-scale is shorter, due to small magnetic Prandtl numbers expected in CV disks \citep{2008Balbus, 2014Potter}. However, as Pm is likely to remain well above $10^{-5}$ \citep{1998Gammie}, the relevant time-scales remain enormous, and only approach the (still large) lower limit of $\sim10^5P_{\rm orb}$ for very cold quiescent states, which are difficult to simulate for other reasons (see previous paragraph). In addition to these enormous times, extremely high resolutions would also be required so that numerical diffusion does not overwhelm the physical extent of these processes. While no existing computational resource would be able to handle such enormous run times, the role of these processes \citep{1998Balbus} has been investigated in high-resolution local studies with artificially high viscosity and magnetic diffusion coefficients at a fixed magnetic Prandtl number \citep[e.g.,][]{2007Fromang, 2008Balbus, 2011Simon}. As a result, a dependency between MRI dynamo efficiency and the Prandtl number has been found for Pm~$\sim1$, which can be of relevance for X-ray binary disks \citep[e.g.,][]{2008BalbusLesaffre, 2014Potter}. In global studies, ensuring a constant non-numerical Pm can be very challenging, especially for CVs, where large time-scale separation due to Pm~$<<1$ is expected, and little flexibility in grid resolution is present due to already high computational cost. These studies typically rely on numerical viscosity to provide grid-scale dissipation of both turbulence and magnetic fields, which results in Pm~$\sim1$ and $\nu,\eta$ that can be difficult to control, especially in context of fast azimuthal flow and mesh refinement. As a result, large-scale magnetic fields are expected to be over-represented in these models, as high Pm promotes efficiency of the MRI dynamo \citep{2001Brandenburg, 2004Schekochihin, 2008Balbus}. With that in mind, however, the \textit{global} structure and behavior of these models should be reflected properly, as argued by \cite{1994Balbus} and shown by \cite{2012Sorathia}, and can be very informative of accretion physics -- as long as one remembers that such models are not proper tools to study the small-scale statistics of MHD turbulence \citep{2011Hawley, 2012Sorathia, 2013Hawley}.

In light of the challenges outlined above, all global (and many local) models of CVs performed to date, including this work, are, by necessity, idealizations. However, idealized models have proven to be very effective in building understanding of various physical phenomena, including disk accretion -- with perhaps the most prominent example in the idealized analytical models of \cite{ShakuraSunyaev}, which are joined by many numerical studies we review below.

\subsection{Global hydrodynamical models}

There is a large body of research using idealized global hydrodynamical models of CVs, with a number of successful predictions. Perhaps the most notable result of these studies is the importance of spiral shock angular momentum dissipation as an accretion mechanism in these systems \citep{1979Lin, 1986Sawada}. Properties of these spirals were thoroughly investigated: from the tidal response's dependence on disk Mach number \citep[e.g.,][]{1994Savonije}, through the efficiency of accretion driving \citep[e.g.,][]{2000Blondin, Wendy1}, factors influencing their opening angles \citep[e.g.,][]{2000Makita}, to their occurrence in various environments \citep[e.g.,][]{2002Belvedere, 2008Lanzafame}. Hydrodynamical models also investigated disk-inflow interactions \citep[e.g.,][]{2001Fujiwara, 2001Kunze, 2019Gordon} and oscillations related to global dynamical instabilities \citep{2007Bisikalo, 2008Kley}.

\subsection{Global MHD models}

However, while the hydrodynamical approach has provided the community with many valuable findings, we now know that the MRI \citep{1992HawleyBalbus} is one of the most important drivers of accretion in semi-detached binaries. Thus, there is a need for idealized global models of accretion disks with the MRI \textit{self-consistently} controlled within an MHD framework, as opposed to its sub-grid treatment via $\alpha$-prescriptions. Such models, despite their limitations discussed in Sect.~\ref{sect:intro:limitations}, are crucial to investigate the interaction between MRI turbulence and global disk structure.

The first global MHD simulations of CVs were performed by \cite{Wendy1,Wendy2}, who used a 3D unstratified setup with realistic treatment of Roche-lobe overflow. They found spiral shocks and MRI to play comparable roles in driving accretion at Mach numbers of $\sim10$ and disk plasma~$\beta\sim400$ (with $\beta$ defined as the ratio of gas pressure to magnetic pressure), with higher magnetization or Mach numbers causing MRI to dominate. They also reported an interplay between these two mechanisms, as more vigorous MRI turbulence was seen to enhance accretion through spiral shocks. Observation of these interactions highlights the potential of idealized global MHD models of accretion disks in semi-detached binaries for improving our understanding of these systems.

Here, we extend these models with vertical stratification. This allows us to probe the vertical structure of accretion disks and compare it with expectations from startified shearing box simulations \citep[e.g.,][]{1995Brandenburg, 1996Stone, 2013Fromang, 2016Salvesen}. We also verify the findings of \cite{Wendy1, Wendy2} regarding spiral shock and MRI accretion in these more physically-consistent, albeit still idealized, conditions. Ultimately, our goal is to include realistic radiative cooling and bridge the global MHD models with local studies using more accurate thermodynamics -- these extensions will be the topic of future papers.

The paper is structured as follows. In Sect.~\ref{sect:model_parameters}, we describe basic parameters of our two models. Sect.~\ref{sect:methods} describes our numerical setup in \texttt{Athena++}. In Sect.~\ref{sect:results}, we report and discuss our results, focusing on global dynamics and comparison with local studies. We shortly discuss possible observational features in Sect.~\ref{sect:results:obs}. Finally, our findings are summarized in Sect.~\ref{sect:conclusions}.
  
\section{Model parameters}\label{sect:model_parameters}
  
 To facilitate comparison with \cite{Wendy1, Wendy2}, we adapt a model mirroring their unstratified setup. We consider a system with mass ratio of $q = M_2/M_1 = 0.3$. The equations we solve are defined in dimensionless form, where $GM_1+GM_2 = 1$. The unit of length is the binary separation $a = 1$, making the binary orbit $P_{\rm orb} = 2\pi$. We investigate two models of accretion disks, with the Mach number at the inner edge of the grid $M_{\rm in}$ equal to~$5$ and~$10$ (corresponding to disk aspect ratios of $H/R\sim1/M_{\rm in}=0.2$ and $0.1$). Henceforth, we will refer to these models as ``Mach~5'' and ``Mach~10''.
 
 

\section{Methods}\label{sect:methods}

  \subsection{\texttt{Athena++}}\label{sect:methods:athena}
  
  Our simulations were performed using the finite-volume 3D~MHD code \texttt{Athena++}\footnote{The public version of \texttt{Athena++} is available at \url{https://princetonuniversity.github.io/athena/}.} \citep{2020StoneAthenaPP}. It is a higher-order Godunov scheme using constrained-transport staggered mesh approach to enforce the zero magnetic divergence constraint \citep{GardinerStone2005, GardinerStone2008}. \texttt{Athena++} includes a number of features tailored towards global MHD simulations, such as flexible grid structure and adaptive mesh refinement. The code has been extensively tested and benchmarked on parallel systems, showing excellent performance and scaling, with well over 80\% parallel efficiency on over half a million threads.
  
  \cadded{\texttt{Athena++} has been designed to be highly modular, with a number of wrappers built in to allow for user modification. Most of the physical features of our models were implemented as user-defined source terms. These include: source terms due to gravity of the binary and rotating frame of reference (Sect.~\ref{sect:methods:mhd_equations}), temperature floor and ceiling (Sect.~\ref{sect:methods:eos}), velocity ceiling (Sect.~\ref{sect:methods:floors:velocity}), and the Alfv\'en speed ceiling (Sect.~\ref{sect:methods:floors:alfven}). Hydrodynamics of our boundary condidions (Sect.~\ref{sect:methods:bvals}) were set with user-defined boundary functions, while initial conditions (Sect.~\ref{sect:methods:init}) were given within a ``problem generator''. Only the EMF boundary and initial conditions (Sect.~\ref{sect:methods:bvals:bvals_L1} and \ref{sect:methods:init}) required base code of \texttt{Athena++} -- the EMF update step -- to be edited.}
  
  \subsection{Equations solved}\label{sect:methods:mhd_equations}
  
  To evolve our models, we solve the equations of ideal magnetohydrodynamics (MHD) in a frame of reference co-rotating with the binary \citep[c.f.,][]{GardinerStone2008, WendyThesis}:
  
  \beq \frac{\partial\rho}{\partial t} + \nablab \cdotb \left(\rho\vec{v}\right) = 0, \label{eq:mhd_density}\eeq
  \beq\begin{array}{rl} \displaystyle \frac{\partial(\rho\vec{v})}{\partial t} + \nablab \cdotb \Big(\rho\vec{v}\vec{v} & - \vec{B}\vec{B} + P^*\tensor{I}\Big)\\
   = & -\rho\nablab\Phi_{\rm tot} + \vec{F}_{\rm cf} + \vec{F}_{\rm Cori}, \label{eq:mhd_momentum}\end{array}\eeq
  \beq \frac{\partial E}{\partial t} + \nablab \cdotb \Big((E+P^*)\vec{v}-\vec{B}(\vec{B}\cdotb\vec{v})\Big) = 0, \label{eq:mhd_energy}\eeq
  \beq \frac{\partial \vec{B}}{\partial t} - \nablab \cross \left(\vec{v}\cross\vec{B}\right) = 0, \label{eq:bfield}\eeq
  
  where $P^*=P+B^2/2$ and $E$ is the total energy density:
  \beq\begin{array}{rl}\displaystyle E = \frac{P}{\gamma-1} + \frac{1}{2}\rho \Big(\vec{v}-\vec{\Omega}\cross(\vec{r} - & \vec{r}_{\rm bary})\Big)^2\\
  \displaystyle+ & \displaystyle\frac{B^2}{2} + \rho\Phi_{\rm tot}. \end{array}\label{eq:mhd_energy_def}\eeq
  $\Phi_{\rm tot}$ is the total gravitational potential of the binary:
  \beq \Phi_{\rm tot} = -\frac{GM_1}{r}-\frac{GM_2}{|\vec{r}-\vec{r}_2|}, \eeq
  where $\vec{r}_2 = 1\vec{e}_x$ is the location of the binary companion. Frame rotation around the barycenter is included through two apparent forces in eq.~(\ref{eq:mhd_momentum}):
  \beq \vec{F}_{\rm cf} = \rho\vec{\Omega}\cross\left(\vec{\Omega}\cross(\vec{r}-\vec{r}_{\rm bary})\right), \eeq
  \beq \vec{F}_{\rm Cori} = -2\rho\vec{\Omega}\cross\vec{v}, \eeq
  where $\vec{\Omega}=1\vec{e}_z$ is the frame rotation rate and $\vec{r}_{\rm bary}\sim0.23\vec{e}_x$ is the location of the barycenter. The remaining symbols have their usual meaning.
    
  \subsubsection{Equation of state}\label{sect:methods:eos}
  
  The system of equations (\ref{eq:mhd_density})~--~(\ref{eq:bfield}) needs to be closed by an equation of state. \cite{Wendy1, Wendy2}, in their unstratified global models, utilized a fixed locally isothermal temperature profile \cadded{with local pressure set by:
  \beq P=\frac{1}{\gamma}c_s^2(R)\rho, \eeq
  where \creplaced{R}{$R$} is the cylindrical radius and $\gamma$ -- adiabatic index.} The \creplaced{temperature}{sound speed $c_s$ (temperature)} at each annulus of the disk was given by the \cite{ShakuraSunyaev} model for a gas-pressure-supported disk with opacity dominated by free-free processes:
  \beq c_s \propto R^{-3/8}, \label{eq:methods:eos:locIsoth}\eeq
  where normalization is set by selecting the Mach number $M_{\rm in}$ at the inner edge of the grid:
  \beq c_s(R_{\rm in}) = \frac{1}{M_{\rm in}}\sqrt{\frac{GM_1}{R_{\rm in}}}. \label{eq:methods:eos:locIsoth2}\eeq
  
  In our stratified models, we find that fixing the temperature profile makes the inflow very hot and pressurized, causing the injected gas to expand in all directions from the L1 zone instead of forming an inflowing stream. To allow the inflow to remain cool until it is shock-heated on impact at the accretion disk, we turn the temperature profile of eq.~(\ref{eq:methods:eos:locIsoth}) into a temperature (pressure) ceiling:
  \beq P_{\rm ceil} = \frac{c_{s}^2(r_{\rm in})}{\gamma}\left(\frac{r_{\rm in}}{r}\right)^{3/4}\rho, \eeq
  where $\rho$ is the local density and the gas is otherwise treated as adiabatic. \cadded{An adiabatic index of $\gamma=1.1$ was chosen to enable direct comparison with numerical models from the literature \citep[e.g.,][]{Wendy1}, which are typically either (locally) isothermal or fully-adiabatic. For the latter, low values of $\gamma$ were needed to prevent disks from becoming too hot. Here, low $\gamma$ prevents our disks from adiabatically cooling too far from our \cite{ShakuraSunyaev} ceiling. Even so, some over-cooling was seen in our test runs with temperature ceiling only. As such cooling quenches MRI (by preventing it from being resolved)}, it has proven necessary to also set a floor on the temperature (pressure) profile:
  \beq P_{\rm floor} = \left(1-f\frac{r-r_{\rm in}}{r_{\rm out}-r_{\rm in}}\right)^k\times P_{\rm ceil}, \eeq
  where $k=2, f=0.95$ (Mach~5) or $k=1, f=0.80$ (Mach~10). As a result, the gas is kept hot and close to the \cite{ShakuraSunyaev} model at low radii, while it is allowed to remain cool at large radii.
  
  \cadded{This simple treatment, with a temperature floor and ceiling limiting an otherwise adiabatic gas, should be sufficient for all processes that depend on the average local temperature (expected to follow \citealt{ShakuraSunyaev}). However, any effects related to small-scale, non-axisymmetric, or transient heating -- e.g., a ``hot spot'' or thermal structure of spiral shocks -- will not be captured. While perhaps of secondary importance to global dynamics, these features are relevant for comparison with observations (see Sect.~\ref{sect:results:obs}). The region most dynamically affected by our thermodynamics treatment is the inflow's impact point, where our implementation artificially enforces a fast-cooling scenario of \cite{1998Armitage} (see Sect.~\ref{sect:results:appearance}). A future version of our models will address these issues with radiative cooling (see Sect.~\ref{sect:conclusions}). This will allow the global dynamical model presented in this work to couple with self-consistent disk thermodynamics.}
  
  \subsection{Simulated domain and mesh layout}\label{sect:methods:mesh}
  
  We represent our models on a spherical-polar mesh with $r\in[0.05,0.62]$, limited by $\sim10$~typical WD radii and the location of the L1~point. It covers a full $4\pi$ of solid angle ($\theta\in[0,\pi]$, $\phi\in[0,2\pi]$) and uses ``polar'' boundary conditions of \texttt{Athena++} to allow for free movement of gas and magnetic fields over both poles of the grid \citep[see][]{AthenaPP}.
  
  The base grid contains 32~cells in each direction, organized in mesh-blocks of $16\times8\times16$~cells (in the directions of $r$, $\theta$, and $\phi$, respectively). Adaptive mesh refinement (AMR) with up to 4 (Mach~5) or 5 (Mach~10) levels of refinement is used to capture the disk and accretion stream.  A mesh-block is marked for refinement if any of the following conditions are satisfied:
  \begin{itemize}
      \item it neighbours the L1 inflow zone, i.e., intersects the region: $|\pi/2-\theta|<\theta_{\rm disk}$, $\phi<\theta_{\rm disk}$ or $2\pi-\phi<\theta_{\rm disk}$, $r=r_{\rm max}=0.62$;
      \item it is close to the midplane ($|\pi/2-\theta|<\theta_{\rm disk}$) and any of its cells have high enough density $\rho > \rho_{\rm AMR}$;
  \end{itemize}
  where we adapt $\theta_{\rm disk}=0.3$ and $\rho_{\rm AMR}=0.05$. Note that \texttt{Athena++} ensures that neighbouring mesh-blocks have refinement levels differing by at most one level \citep{2020StoneAthenaPP}. While our needs could have been satisfied with static mesh refinement, we find AMR to be more flexible and convenient at a negligible additional cost. Our grid is logarithmically spaced in radius, with cell size ratio of $1.1$ in this direction. As a result, the aspect ratio of cells remain close to $1:1:2$ throughout the grid, with the elongated part directed azimuthally. In both our simulations the accretion disk was always resolved at the highest refinement level and the thermal scale height was resolved by $H/\Delta z\simeq27$ cells. We discuss our MRI resolving power in Sect.~\ref{sect:results:bfield}.
  
  The time~step of our simulations was limited by advection of gas with Keplerian velocity at the midplane, at the inner edge of the grid ($r=0.05$, $\theta=\pi/2$; however, see Sect.~\ref{sect:methods:floors:alfven}). Note that, to conserve numerical resources, we increased the inner grid radius from $0.02$ used by \cite{Wendy1, Wendy2} to $0.05$ here.
  
  \subsection{Floors and ceilings}\label{sect:methods:floors}
  
    \subsubsection{Density floor}\label{sect:methods:floors:density}
    
    Our simulations contain large volumes of void threaded by magnetic fields -- especially in the polar regions. To ensure numerical stability, we find it necessary to use a density floor at $\rho_{\rm floor}=10^{-6}$, much smaller than typical disk densities of $\rho\sim 1$. Note that by using a density floor we inject additional mass in a frame of reference co-rotating with the binary. We find this to be a physically reasonable approach, as some amounts of ambient gas co-rotating with the disk may be present in astrophysical systems, while a reservoir of mass immobile in the (inertial) LAB frame appears unlikely. Thanks to application of a velocity ceiling (see Sect.~\ref{sect:methods:floors:velocity}, eq.~\ref{eq:mdotfloor}), density floor contributes negligibly to total accretion rates observed in our simulations.
  
    \subsubsection{Velocity ceiling}\label{sect:methods:floors:velocity}
    
    Near the equatorial plane of our models, gas can readily be supported by Keplerian rotation and pressure gradients. However, near the polar regions it can at times be in free-fall, causing a number of numerical issues common in global models of accretion disks. To mitigate them, we adopt the following velocity ceiling.
    
    \cadded{If (and only if) the total velocity at a given cell exceeds the critical value, $v>v_{\rm ceil}=0.5$, a smooth switch $s_v$ is calculated. It applies the velocity ceiling only to low density regions, $\rho \lesssim \rho_v = 10^{-5}$ (to avoid affecting disk dynamics), with a smooth transition in velocity:}
    \beq\begin{array}{rl}
        s_v =& \left(\arctan \left(\left(1-\frac{\rho}{\rho_v}\right)\times8\pi\right)/\pi+\frac{1}{2}\right) \nonumber\\
        &\times\left(\arctan\left(\left(\frac{v}{v_{\rm ceil}}-1.2\right)\times8\pi\right)/\pi+\frac{1}{2}\right).\end{array}
    \label{eq:smoothing_velocity} \eeq
    \cadded{Factors of $8\pi$ and $1.2$ are chosen to balance smooth behavior with well-defined application boundaries. $s_v$ is then} used to compute the adjusted velocity:
    \beq v_{\rm new} = s_vv_{\rm ceil} + (1-s_v)v, \eeq
    \creplaced{Note that $s_v$ limits velocity ceiling application to low-density regions with $\rho \lesssim \rho_v = 10^{-5}$ (first line of eq.~\ref{eq:smoothing_velocity}), preventing the velocity ceiling from affecting disk dynamics. Only radial and poloidal components of the velocity are adjusted with $v_{\rm new}$}{which modifies only the radial and poloidal components of cell velocity}:
    \begin{eqnarray}
        v_{r, \rm new} &= (v_{\rm new}/v)\times v_r, \\
        v_{\theta, \rm new} &= (v_{\rm new}/v)\times v_{\theta}.
    \end{eqnarray}
    As all source terms applied to momenta of the system, the velocity ceiling also includes an adjustment to the total energy density (eq.~\ref{eq:mhd_energy_def}):
    \beq\begin{array}{rl} \Delta E = \frac{1}{2\rho} & \left((\vec{M})^2 - (\vec{M}-\Delta\vec{M})^2\right) \\
    & = \frac{1}{2\rho}\left(2\vec{M}\cdotb\Delta\vec{M} - (\Delta\vec{M})^2\right), \end{array}\label{eq:Eupdate}\eeq
    where $\vec{M}$ is the momentum density \textit{after} velocity ceiling application and $\Delta\vec{M}$ is the vector by which it is changed.
    
    In addition to improving stability, the velocity ceiling also ensures that the accretion rate of material at the density floor from polar regions does not dominate our results. We can approximate the density floor accretion rate as:
    \beq \dot{M}_{\rm floor} \simeq 4\pi r_{\rm out}^2\times \rho_{\rm floor}v_{\rm ceil} \simeq 2.42\times 10^{-6}\textrm{ sim.u.}, \label{eq:mdotfloor}\eeq
    which remains well below the observed disk accretion rates, reported in Fig.~\ref{fig:mdot} and Sect.~\ref{sect:results:variability}.
    
    \cadded{While the velocity ceiling does affect magnetic fields in the polar regions of our grid, we find that majority of mass and magnetic energy of our models resides in regions of parameter space unaffected by these changes. We confirmed this by inspecting 2D histograms of magnetic energy density and gas density, weighted by these two quantities.}
    
    \subsubsection{Alfv\'en speed ceiling}\label{sect:methods:floors:alfven}
    
    In our Mach~10 run, shock structures within the disk sometimes produced very small regions (few cells each) of density floor permeated by strong magnetic fields. These resulted in small time~steps, halting the simulation. To prevent this, we imposed an Alfv\'en speed ceiling. Its value was set at the Keplerian speed of the inner grid edge (which normally sets the time~step), $v_{A\rm, ceil} = \sqrt{GM_1/r_{\rm in}} \simeq 3.9$. Whenever it was crossed, local density was increased for $v_A$ to match that value. We have measured the rate of mass injection by this modification to density floor, and it was at most $10^{-3}$ of the observed physical accretion rate.
  
  \subsection{Boundary conditions}\label{sect:methods:bvals}
  
  Outside of the L1 zone, the inner and outer radial boundary conditions are set as free-outflow, no-inflow (``diode'' boundary conditions). If the local radial velocity is directed outside of the simulated grid, both the cell-centered quantities and the edge-centered EMFs (see \citealt{GardinerStone2008}) are copied to the ghost cells, allowing for free outflow. If the radial velocity is directed into the grid, reflecting boundary conditions are used for cell-centered values and EMFs in the ghost cells are set to $0$. The boundary conditions in the $\phi$~direction are set as periodic. We use special ``polar'' boundary conditions \citep{AthenaPP} for the boundaries in the $\theta$~direction. These allow us to accurately represent motion of gas and magnetic fields even as they pass through the poles of the grid.
  
  \subsubsection{Roche-lobe overflow boundary conditions}\label{sect:methods:bvals:bvals_L1}
  
  In order to simulate the inflow of magnetized gas through the L1 point of the binary system, we designate a special ``L1 zone'' in the outer radial boundary. It is defined as a circular region at an angle of at most $0.10$~rad ($\sim6\degree$, Mach~5) or $0.05$~rad ($\sim3\degree$, Mach~10) to the binary axis as seen from the grid center.
  
  For L1 zone ghost cells, the density is set to $1$ (Mach~5) or $4$ (Mach~10), and a small negative radial velocity of $v_r=-0.01$ is set, while $v_{\theta}$ and $v_{\phi}$ are kept at $0$. \creplaced{We note that the actual mass inflow rate}{While this boundary $\rho v_r$ gives us an indirect handle on the mass inflow rate, the actual value of inflow $\dot{M}$} depends on \cadded{active} grid conditions \cadded{surrounding the L1 zone (mainly gas and ram pressure)}, and is best measured empirically from the simulation outputs as $\dot{M}$ at large radii. \cadded{Systematically larger inflow $\rho v_r$ would lead to a denser disk with the same aspect ratio, as long the disk reaches our temperature ceiling (where $H/R\sim1/M_{\rm in}$). However, any potential short-timescale fluctuations in inflow $\rho v_r$ would likely be quickly ``forgotten'' by the inflow stream, given the level of variability caused by interaction with the gas surrounding our accretion disks (see Sect.~\ref{sect:results:variability}). Such inflow variability can drive some level of turbulence within the disk, but it is likely quickly overridden by MRI effects once magnetic fields are taken into account (as observed for corresponding hydrodynamical and MHD models of \citealt{Wendy1}).}
  
  \creplaced{The}{Inflowing} gas is injected at the local temperature floor (see Sect.~\ref{sect:methods:eos}) with the sound speed corresponding to $0.020$ and $0.174$ of the equatorial sound speed at $r_{\rm in}$ for Mach~5 and Mach~10, respectively.
  
  The inflow also contains magnetic field. We take a zero-net-flux approach with alternating-polarity vertical magnetic loops traveling inside the inflow (the loop axes are in the $\phi$~direction)\footnote{This orientation prevents numerical reconnection as the loops are sheared down to a few radial rows of cells when the inflow reaches the forming accretion disk.}. We opt to set the magnetic fields of L1 ghost zone by modifying the EMFs, allowing the code to calculate corresponding $B$-fields. This ensures $\nablab\cdotb\vec{B}=0$ throughout the grid at all times. The inflow EMF values are set as follows.
  \begin{enumerate}
      \item For a single (circular) row of cells surrounding the L1 zone, EMF values are always set to 0. This prevents magnetic fields from ``spreading'' along the outer radial boundary of the grid and keeps them confined to the inflow.
      \item Within the L1 zone, the following condition is used:
      \beq \vec{E} = \pm\left( \sqrt{\frac{2P_{\rm infl}}{\beta_{\rm infl}}} \frac{ x_{\rm loop} }{ d } v_{\rm infl} \right)\vec{e_{\phi}}, \label{eq:emf}\eeq
      with the sign alternating between loops. $P_{\rm infl}$ denotes the inflow pressure, $\beta_{\rm infl}$ -- minimal plasma $\beta$ within the loop, and $x_{\rm loop}$ describes how far the current loop would have advected into the grid if it were moving at $qv_{\rm infl}$:
      \beq x_{\rm loop} = \left(-qv_{\rm infl}t \textrm{ mod } 2r_{\rm loop}\right) - r_{\rm loop}, \eeq
      \beq d = \sqrt{\left(r\cos\theta\right)^2 + x_{\rm loop}^2} + 10^{-6}. \eeq
      The ``squeezing factor'' $q=2\pi$ is used to prevent loops from being excessively elongated. The loop radius is set to match the L1 zone radius.
  \end{enumerate}
  
  The \cremoved{optimum} strength of the injected magnetic field \cadded{depends on interactions with the active grid and thus a value} that produced dynamically important (MRI unstable) magnetic field in the disk \cadded{needed to be} set by trial and error. In the data presented here, mass-weighted averages of plasma~$\beta$ over the L1 vicinity are $66.5$ (Mach~5) and $343$ (Mach~10). These inflows are additionally heated by our temperature floor, which increases the effective plasma~$\beta$ of the gas reaching the disk by another factor of $(c_{s,\rm disk}/c_{s,\rm infl})^2$ to $\sim345$ and $\sim487$ for Mach~5 and Mach~10, respectively. These final estimates are consistent with typical values used in similar studies \cite[e.g.,][]{Wendy1, Wendy2}. \cadded{We note that, in absence of an MRI dynamo, magnetic fields at the disk midplane would be affected by numerical reconnection due to Keplerian shear and thus decay noticeably within the disk (this is not observed).}
  
  \subsection{Initial conditions and discarded transients}\label{sect:methods:init}
  
  Both simulations are initialized with a thin magnetized low-density Keplerian disk surrounded by density floor. For $r < 0.3$, the following conditions are used:
  \beq \rho = \rho_{\rm init}\times\xi(r,z)+ \rho_{\rm floor}, \eeq
  \beq v = \frac{\sqrt{GM_1}R}{r^{3/2}} - \Omega R, \eeq
  \beq\begin{array}{rl}
      \xi(r,&z) = \left(\frac{1}{2}-\arctan\left(16\pi\left(r-r_{\rm init}\right)\right)/\pi\right) \\
      &\times\left(\arctan\left(\frac{z+H_{\rm init}}{s}\right) - \arctan\left(\frac{z-H_{\rm init}}{s}\right)\right)/\pi,
  \end{array}\eeq
  where $r_{\rm init} = 0.2$, $\rho_{\rm init} = 0.1$, $H_{\rm init} = 0.05$, $s=0.005$ is the smoothing parameter, $R$ and $r$ are the cylindrical and spherical-polar radius, respectively, and $\Omega=1$ is the rotation speed of the frame of reference. The disk is initialized as magnetized with a single magnetic loop defined using:
  \beq \vec{E} = vB_{\rm init} \times \frac{\cos\left(\frac{z\pi}{2H_{\rm init}}\right)\sin(\phi)}{r^2\sin\theta}  \times\xi(r,z)\vec{e}_r, \eeq
  where $B_{\rm init} = 1$. At $r > 0.3$, void conditions are set with $\rho=\rho_{\rm floor}$, $v=0$, and $\vec{E} = 0$. The pressure is set according to eqs.~(\ref{eq:methods:eos:locIsoth}),~(\ref{eq:methods:eos:locIsoth2}).
  
  This initial state was first evolved with density floor and velocity ceiling's critical density of $10^{-4}$ and $10^{-3}$, respectively. After $4$ (Mach~5) or $15$ (Mach~10) binary periods, well-evolved accretion disks were present. We then set $\rho_{\rm floor}$ and $\rho_v$ to $10^{-6}$ and $10^{-5}$, ensuring that floor accretion is no longer able to affect the accretion rate (see Sect.~\ref{sect:methods:floors}). In order to avoid the influence of transients associated with establishing the accretion flow using this procedure, the initial $\sim5P_{\rm orb}$ of subsequent evolution was ignored, with the remaining $\sim9P_{\rm orb}$ (Mach~5) and $\sim11P_{\rm orb}$ (Mach~10) used to perform the analysis presented in this work.

\section{Results and Discussion}\label{sect:results}

  As the lengthy discussion of the methods given in the previous section implies, self-consistent modeling of accretion in a global model of a close binary is difficult. Even with advances in computational methods and infrastructure (such as AMR in curvilinear coordinates), we remain very much limited by computational constraints. Nonetheless, even simplified models can provide valuable insights into the behavior of real systems and guide future enhancements to the models. Providing these insights is our goal as we report the findings from our stratified global MHD simulations of an accretion disk with Roche-lobe overflow.

  \subsection{General description of the flow}\label{sect:results:genDescr}
  
    \begin{figure*}
        \centering
        \begin{tabular}{cc}
            \includegraphics[scale=0.225]{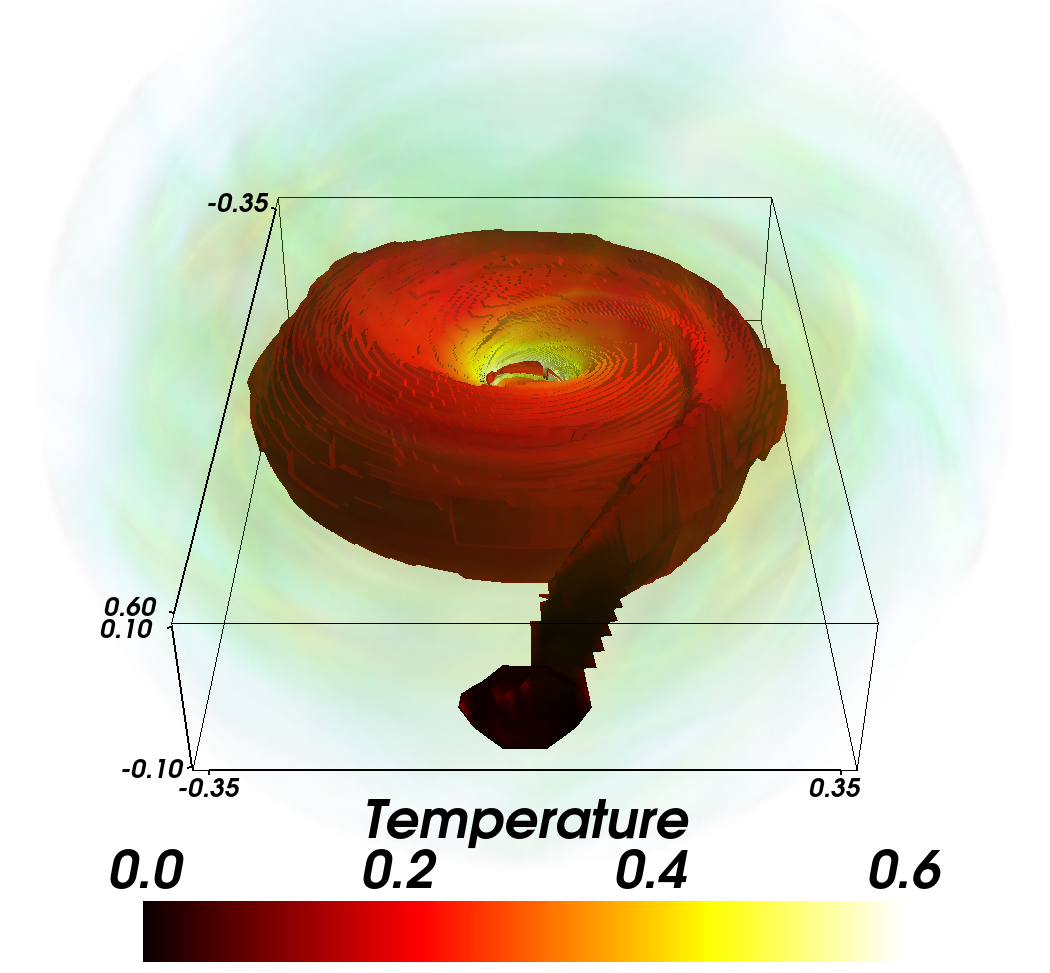} & \includegraphics[scale=0.225]{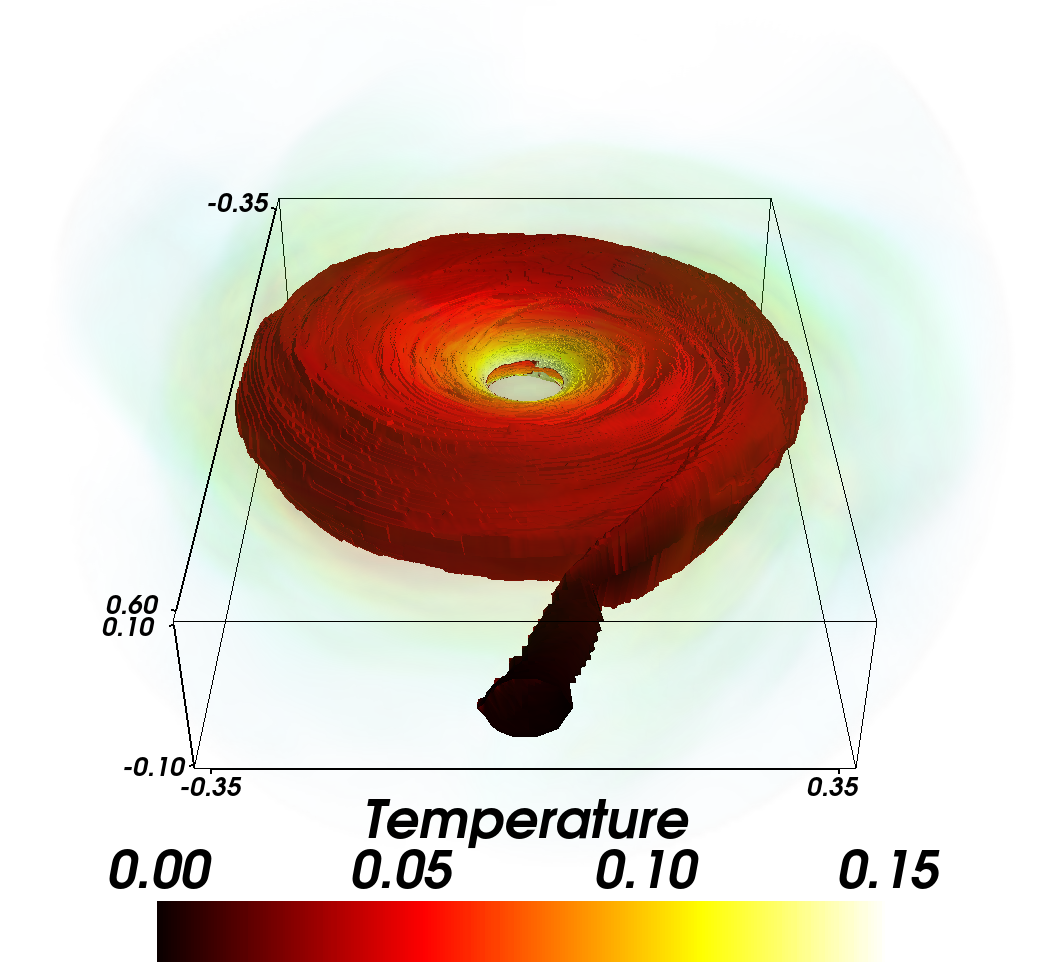} \\
        \end{tabular}
        \caption{Renderings of density isocontours for Mach~5 at $t=7.6P_{\rm orb}$ (left) and Mach~10 at $t=9.4P_{\rm orb}$ (right) for $\rho=0.015$~sim.u., corresponding to $0.30$ and $0.03$ of the central disk density, respectively. Surface color denotes disk temperature at a given point in sim.u. and volume rendering surrounding the disk shows magnetic field structure (see text for a detailed description). Bounding box with $x$, $y$, and $z$ extents labelled is shown for scale. \cadded{For a more detailed view of the disk structure, we refer the Reader to simulation slices shown in Figs.~\ref{fig:snapshot_rho} and~\ref{fig:snapshot_beta}.}}
        \label{fig:3D}
    \end{figure*}
  
  We plot 3D~renderings of our two models in Fig.~\ref{fig:3D}. The data are taken from snapshots of Mach~5 (left panel) and Mach~10 (right panel) at $t=7.6P_{\rm orb}$ and $t=9.4P_{\rm orb}$ (after the discarded transients, Sect.~\ref{sect:methods:init}), respectively. Surface plot corresponds to a density isocontour at $0.30$ (Mach~5) and $0.03$ (Mach~10) of the central disk density (as measured at $r=0.08$~sim.u., \creplaced{configuration}{phase}~A for Mach~10). The disk density isosurfaces readily show signs of spiral structure, which we discuss in detail in Sect.~\ref{sect:results:appearance} and~\ref{sect:results:spiral}. Surface color denotes disk temperature at a given point in sim.u. Note that the factor of 2 difference in $M_{\rm in}$ between the two models results in a factor of 4 difference in temperature. The surface temperature confirms that the inflow remains cold until it is shock-heated within the disk into \cite{ShakuraSunyaev} profile (our temperature ceiling), as intended. Volume rendering of magnetic field strength (seen as light colors surrounding the disk) visualizes location and shape of magnetic structures. While in Mach~5 strong magnetic fields are seen to fill the entire surroundings of the disk, in Mach~10 they are mostly limited to regions just above the disk surface. We discuss the magnetic field structure in detail in Sect.~\ref{sect:results:bfield}.
  
  A number of authors have considered the inflow -- accretion disk interactions by means of analytical considerations and with the use of numerical (radiative) hydrodynamics \citep[e.g.,][]{1976Lubow, 1986Livio, 1987Frank, 1989Lubow, 2001Kunze, 2019Gordon}. \cite{1998Armitage} found that if cooling in the system is efficient, the inflow stream can reflect off the rim of the accretion disk leaving a bulge in the disk downstream of the impact point. In the case of inefficient cooling, the stream is seen to overflow the accretion disk, with smaller streams continuing to slide over the disk surface following near-ballistic trajectories \citep[e.g.,][]{2001Kunze}. Our models do not match either of these scenarios exactly, as cooling is applied only at the temperature ceiling (at which point it is nearly infinitely efficient). However, qualitatively, they resemble the efficient cooling scenario of \cite{1998Armitage} -- a small elongated downstream bulge (clearly related to the underlying spiral structure) is indeed visible at the inflow impact point (especially for Mach~5, see Fig.~\ref{fig:3D}) and some reflection of the inflowing material towards larger radii is seen in the density snapshots (see Sect.~\ref{sect:results:appearance}). \cadded{However,} we see no evidence of stream overflow \cadded{(i.e., passing over the impact point to continue as smaller streams ``sliding'' over the disk surface)}. We note that this behavior is likely influenced by the fact that accretion disks in our models operate close to the temperature ceiling. While there is an elongated shock structure at the inflow impact point, a ``hot spot'' is absent from our models, where the gas instead heats up gradually. This is not unexpected, considering our constraints on temperature (Sect.~\ref{sect:methods:eos}). We intend to investigate the behavior of disk-inflow interaction, including the ``hot spot'' and overflow streams, more closely in our future work with more realistic temperature treatment (see Sect.~\ref{sect:conclusions}).
  
    \subsubsection{Variability}\label{sect:results:variability}
    
        \begin{figure*}
            \centering
            \includegraphics{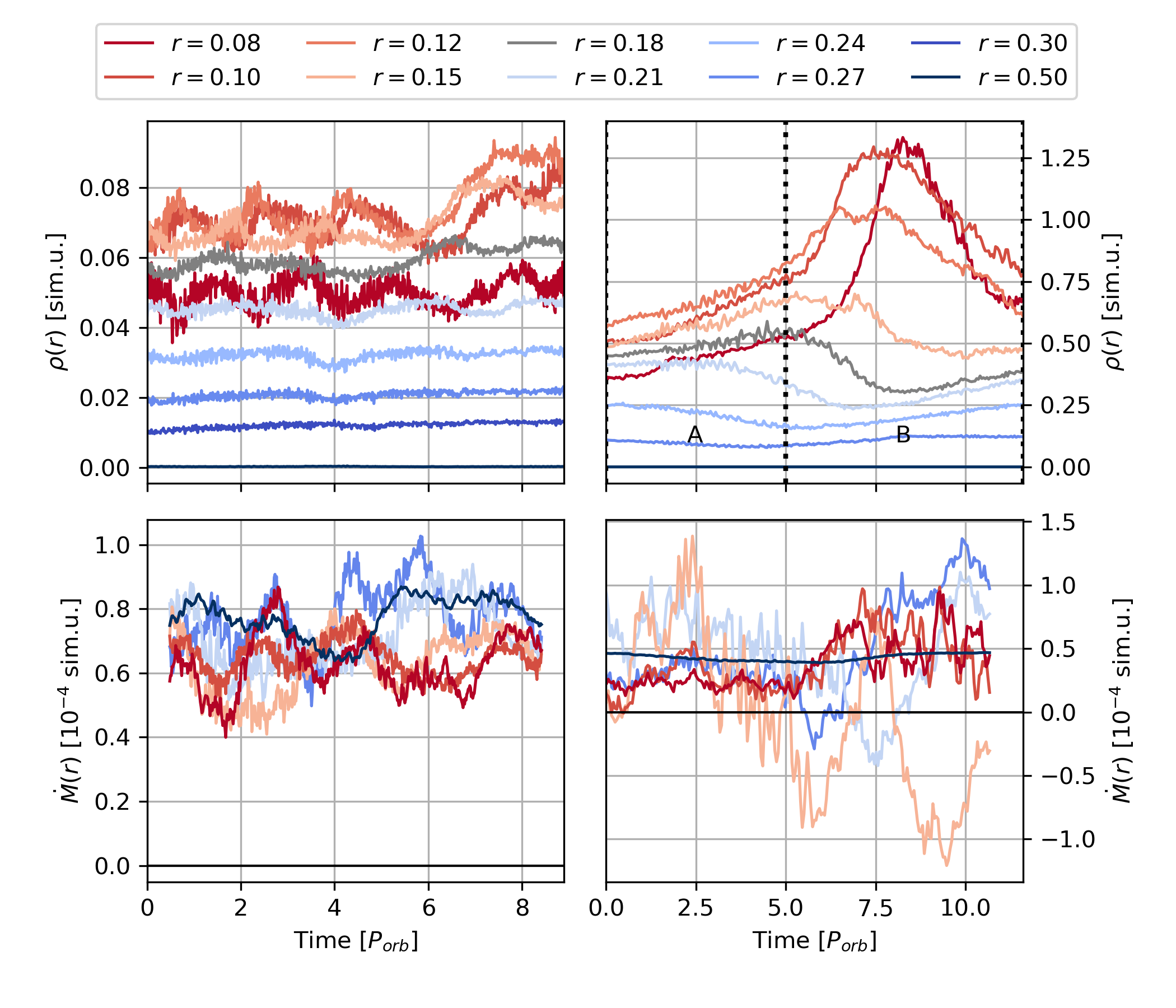}
            \caption{Evolution of density structure and accretion rate for Mach~5 (left column) and Mach~10 (right column). Top: average disk density at a given radius (indicated by the curve color). Bottom: instantaneous accretion rate at a given radius. \cadded{Note that all curves are of the same thickness, their apparent widening is caused by strong variability.} Vertical dashed line in the top right plot shows the boundary between the disk \creplaced{configuration}{phase}s A and B, as discussed in the text.}
            \label{fig:mdot}
        \end{figure*}
    
    Accretion disks of semi-detached binaries exhibit variability at a multitude of timescales and amplitudes: from DN outbursts \citep[e.g.,][]{1971Smak, 1974Osaki, 1981Meyer, 2001Lasota}, through various types of QPOs \citep{2003Warner}, to rapid flickering \citep[e.g.,][]{1992Bruch, 2001Sokoloski, 2004Woudt}. Nonetheless, it is commonly assumed that, at most times, a sufficiently long time average of accretion disk observables is well described by a steady-state model. Whether or not accretion disks truly reach such quasi-stationary states (and over how long a time) remains a viable question.
    
    In Fig.~\ref{fig:mdot}, we show two metrics we use to evaluate whether a steady state is present in our models. First, we consider how radial density profiles of the disk change with time. We use density averaged over $\theta\in[\pi/2-\theta_H, \pi/2+\theta_H]$, $\phi\in[0,2\pi]$ at a number of radii, where $\theta_H=1/M_{\rm in}$ roughly corresponds to the local thermal scale height. These time series are then boxcar-averaged over a single binary period to remove high-frequency effects. The resulting density evolution plots are shown in the top panels of Fig.~\ref{fig:mdot}, color-coded by radius. We also calculate instantaneous accretion rates through the selected radii. These are the mass flow rates integrated over the full sphere at given radii, with time series similarly boxcar-averaged over one orbital period. We show these data in the bottom panels of Fig.~\ref{fig:mdot}.
    
    For comparison, we calculate that the dynamical time-scale in our disks ranges from $2\times10^{-3}P_{\rm orb}$ at $r_{\rm in}$ to $3\times10^{-2}P_{\rm orb}$ at $r=0.3$, while the effective viscous time-scale:
    \beq \tau_{\rm visc} = \frac{R^2}{\alpha c_sH} \sim \frac{R_{\rm in}^{1/8}M_{\rm in}^2}{\alpha\sqrt{GM_1}}R^{11/8} \eeq
    is equal to $\sim6P_{\rm orb}$ and $\sim24P_{\rm orb}$ for Mach~5 and Mach~10, respectively, assuming $\alpha\sim0.1$ and $R\sim0.3$. Thus, given that our simulations were run for $9P_{\rm orb}$ (Mach~5) and $20P_{\rm orb}$ (Mach~10), respectively, \textit{before} the results shown in Fig.~\ref{fig:mdot} (see Sect.~\ref{sect:methods:init}), in each case our simulations cover roughly a viscous time in the disk. However, given the actual $\alpha$ values we measure in the flow are smaller (Sect.~\ref{sect:results:radial}), it is clear our models must still be considered exploratory. To fully address long term dynamical evolution at high Mach numbers, significantly longer (and many times more expensive) simulations are needed. If such ambitious models are attempted in the future, we hope that this work may provide a guideline for what can be expected.
    
    As evident from Fig.~\ref{fig:mdot}, both of our models show density and accretion rate variability over a wide range of time scales.
    However, diagnostics of Mach~5 oscillate around well-defined average levels, describing a valid stationary state with the average $\dot{M}$ roughly independent of radius. In contrast, Mach~10 exhibits variability at all time scales, including that of the analyzed window itself ($\sim11P_{\rm orb}$). We find that longest time scale variability to be particularly interesting. Up until $t\sim5P_{\rm orb}$ (marked with a vertical dotted line in the top right panel of Fig.~\ref{fig:mdot}), the inner parts of the disk ($r\lesssim0.12$) operate in a quasi-stationary state, at an accretion level lower than that of the inflow (seen as $\dot{M}(r=0.5)$). As a result, gas is accumulated throughout the disk, causing density to increase for $r\lesssim0.2$. We will refer to this episode of low-level quasi-steady accretion as ``\creplaced{configuration}{phase} A''. Around $t=5P_{\rm orb}$, average densities at larger (bluer) radii start to drop, causing a cumulative increase in density in each consecutive (smaller-redder) radius, until the (so increased) density at that radius drops as well. We observe a runaway, where gas from the outer radii travels inwards, gathering mass accumulated in the disk in an outside-in fashion (\creplaced{we refer to this phase as ``configuration}{we refer to this as ``phase}~B''). Once the event passes a given radius, local density regains its initial value and the accumulation of gas begins anew. This is suggestive of a recurring phenomenon, although longer simulations are needed to verify this hypothesis. Alternatively, \creplaced{configuration}{phase} B may be a non-stationary feature, as a quasi-steady state may exist at a later time beyond the simulation time available to us. Even then, however, it may be present in systems adjusting to a recent change in accretion rate either through the inflow stream or in the disk itself (e.g., during an outburst). We continue our discussion of \creplaced{configuration}{phase}s A and B in Sect.~\ref{sect:results:Mach10}, where we look closer at the differences between the two regimes.
    
    \subsubsection{Snapshots of density, spiral structure}\label{sect:results:appearance}
    
        \begin{figure*}
            \centering
            \includegraphics{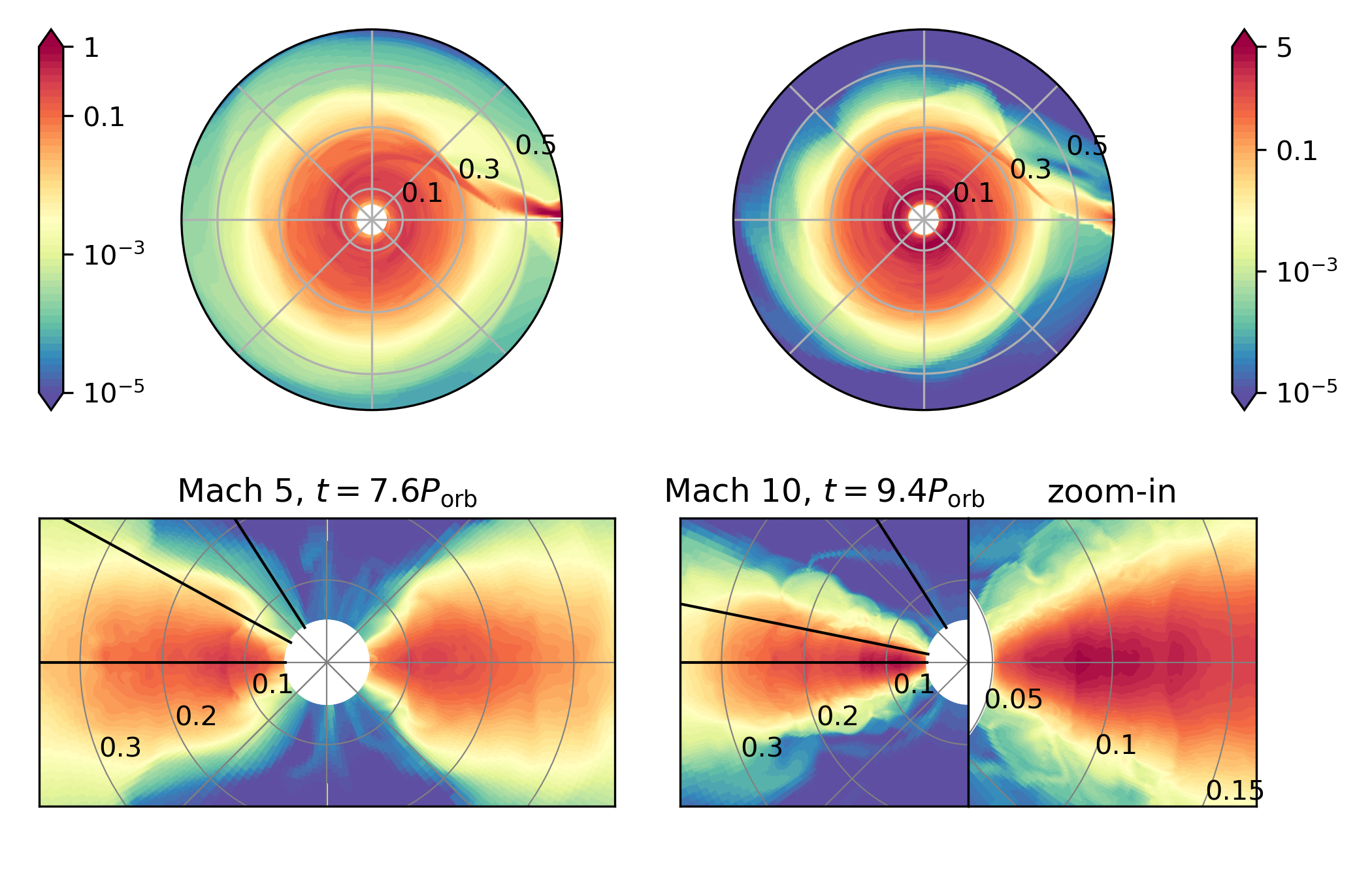}
            \caption{Density snapshots for Mach~5 (left) and Mach~10 (right). Top: full equatorial slices, bottom: poloidal slices of the disk region ($\phi=0,180\degree$ for right and left halves, respectively). Solid black lines in the bottom plots visualize the $\theta$ ranges used to define ``disk body'' and ``corona'' regions (see Sect.~\ref{sect:results:bfield} for details).}
            \label{fig:snapshot_rho}
        \end{figure*}
    
    In Fig.~\ref{fig:snapshot_rho}, we show density snapshots of our models at $t=7.6P_{\rm orb}$ (Mach~5) and $t=9.4P_{\rm orb}$ (Mach~10), depicted using equatorial and poloidal slices. The latter are taken for $\phi=0,180\degree$ for the right and left halves of the plots in the bottom panels, respectively. A zoom-in is used in the $\phi=0\degree$ plot for Mach~10. The solid black lines in the bottom panels indicate polar angle ranges for the disk's ``main body'' \cadded{at an opening angle of $\sim28\degree$ for Mach~5 and $\sim11\degree$ for Mach~10} ($|\pi/2-\theta|\in [0, 2.5H/R]$ and $|\pi/2-\theta|\in [0, 2H/R]$ for Mach~5 and Mach~10, respectively) and ``corona'' \cadded{further extending to $\sim57\degree$ in both cases} ($|\pi/2-\theta|\in [2.5H/R, 5.0H/R]$ and $|\pi/2-\theta|\in [2H/R, 10H/R]$), which we discuss in Sect.~\ref{sect:results:bfield}.
    
    The inflow enters each accretion disk from the right through an elongated shock structure, seamlessly transitioning into one of the disk's spiral arms. These can be seen both as overdensities in the equatorial plots (top panels of Fig.~\ref{fig:snapshot_rho}) and as shock structures in the poloidal slices (bottom panels of Fig.~\ref{fig:snapshot_rho}). Vertically, these spiral shocks are inclined and occasionally broken into multiple parts with different inclinations, as is the case, for instance, for the shock at $r\sim0.12,\phi=180\degree$ in the bottom left panel of Fig.~\ref{fig:snapshot_rho} (Mach~5) or one at $r\sim0.125,\phi=0$ in the bottom right panel of Fig.~\ref{fig:snapshot_rho} (Mach~10). Both radial position of the spiral arms and vertical outlines (inclination as a function of height) of the associated shocks change rapidly. For both Mach~5 and Mach~10, the spiral arms regularly deviate by up to $\sim0.05$~sim.u. from their average positions, and inclination can rapidly change within the bounds of $\sim[-\pi/4,\pi/4]$, with opposite extrema of this range often seen at consecutive time snapshots, $0.01$--$0.05P_{\rm orb}$ apart.
    
    The spiral shocks in the Mach~5 model appear to influence the vertical extent of the disk. A difference in height between pre- and post-shock regions can be seen in the bottom left plot of Fig.~\ref{fig:snapshot_rho}. Aside from these changes, however, the main body of the Mach~5 disk, extending up to $\sim3$ vertical scale heights from the midplane, appears to be well mixed, with little vertical structure.
    
    Spiral structure in semi-detached binaries has been extensively studied in purely hydrodynamical simulations, both using smoothed-particle hydrodynamics \citep[SPH; e.g.,][]{2002Belvedere, 2002LanzafameCostaBelvedere, 2003Lanzafame, 2010Lanzafame} and with grid-based models \citep[e.g.,][]{2000Makita, 2001Fujiwara, Wendy1, 2017Geoffrey, 2017Lukin}. As magnetic fields are dynamically sub-dominant in our models, many of the aspects of spiral shocks seen in our models are similar to ones reported in hydrodynamical framework. The elongated shock structure connecting our inflow to the spiral arms, sometimes dubbed a ``hot line'', is a frequent feature in these studies (e.g., fig.~4 of \citealt{2017Lukin} and its discussion therein; however, see also \citealt{1998Bisikalo}). Dependence of the spiral pitch angle on the disk Mach number has been reported and discussed by a number of authors \citep[e.g.,][]{1987Spruit, 2016Hennebelle, Wendy1} and the spiral pattern we observe in the equatorial plane is similar to ones reported in hydrodynamical studies at similar parameters (e.g., compare our Fig.~\ref{fig:snapshot_rho} with fig.~3 of \citealt{2000Makita} or fig.~14 of \citealt{Wendy1}). Increase in vertical extent of the disk caused by underlying spiral structure has also been previously observed in non-magnetic runs (e.g., figs.~5,7 of \citealt{2000Makita}). Pitch angles of spiral arms in semi-detached binary disks are usually found to follow linear dispersion relations \citep{Wendy1, 2017Geoffrey}, which, as we discuss in detail in Sect.~\ref{sect:results:spiral}, is also found to be true here for the \textit{time-averaged} spiral pattern. However, an apparent difference between hydrodynamical studies and our results lies in the level of variability in spiral shock position and vertical inclination. Spiral patterns in hydrodynamical models of accretion disks are generally described as stable, settling into a well-behaved steady state. While there are indications of inclined shocks (see, e.g., fig.~6 of \citealt{2000Makita}, discussion in \citealt{2001Fujiwara}, fig.~4 in \citealt{2003Lanzafame}), the vertical patterns appear to be symmetric with respect to the disk midplane. This stands in contrast with our models, where position and inclination of shocks can change significantly on time scales much smaller than a binary period, and the spiral shocks are often found to be asymmetric with respect to the disk midplane. This is likely caused by interactions with the underlying MRI turbulence. \cite{Wendy1, Wendy2}, in their unstratified MHD simulations, also note some disruption of the spiral pattern by the MRI turbulence, although they do not comment on its variability. Position and inclination of spiral shocks may be more significantly affected in our stratified models, where an additional (vertical) degree of freedom is introduced (an effect similar to what is seen as distortion of spiral waves in 3D models of \citealt{2001Fujiwara}). Alternatively, variability of the spiral shock morphology may be related to high resolution of our models, which resolves a wide range of Kelvin-Helmholtz instabilities in the flow -- an effect similar to what is seen in the convergence study of \cite{Wendy1} (see their fig.~14).
    
    For Mach~10, the densest regions ($\rho\gtrsim1$) do not appear to be sensitive to the spiral structure. The disk surface is almost conical, at $\sim2$ vertical scale heights from the midplane. Above it, a low-density strongly magnetized ``corona'' is present (see Sect.~\ref{sect:results:bfield}). In density snapshots, shell-like overdensities are produced by parcels of magnetic field buoyantly rising from the disk, as they drive some of the main-body gas over the disk surface. 
    
    Interestingly, some of the inflowing gas in our Mach~10 model is initially reflected away at radially supersonic speeds, which results in a series of plumes orbiting around the disk at $r\gtrsim0.3, \phi\lesssim\pi$ (top right panel of Fig.~\ref{fig:snapshot_rho}, most recent plume at $\phi=3\pi/4, r\sim0.4$), until they settle into the accretion disk at larger $\phi$. These plumes follow a self-regulating cycle. While a plume is present above the inflow stream's impact point, the reflected gas is shocked not only by the accretion disk's rim (which causes the initial reflection), but also the plume of previously reflected gas -- thus entrapping the inflow and stopping formation of a new plume until the impact point is clear of reflected gas (a process resembling interactions with a circumbinary envelope investigated by \citealt{1998Bisikalo}). This cycle causes plume formation to occur at regular intervals with each plume $\sim\pi/4$ apart in $\phi$. As discussed in Sect.~\ref{sect:results:appearance}, inflow stream reflection has been predicted for efficiently cooling disks by \cite{1998Armitage}, and this scenario appears to be applicable here due to our temperature ceiling being active within the disk region.
    
    In accreting binaries, the disk is influenced by tidal interaction with the binary companion. This results in truncation at a radius $r_{\rm tid}$ where tidally-distorted orbits cross \citep{1977Paczynski, 1990Hirose}. For our adopted $q=0.3$ \citep[e.g.,][]{warner_1995, 1996HarropAllin}:
    \beq r_{\rm tid} \simeq \frac{0.60}{1+q}a \simeq 0.46\textrm{ sim.u.} \eeq
    As can be seen in the top row of Fig.~\ref{fig:snapshot_rho}, the outermost parts of both disks do extend only until that radius. However, a distinct densest part of each disk appears to be enclosed within $r\lesssim0.3$. This may be caused by spiral shocks, which cause gas orbits to be eccentric, increasing chances of orbit crossings, and thus lower the actual outer truncation radius. Radial size of our disk is generally consistent with corresponding models of \cite{Wendy1, Wendy2}, indicating that stratification does not influence this property of the disk significantly.
    
    \subsubsection{Magnetic fields}\label{sect:results:bfield}
        
        \begin{figure*}
            \centering
            \includegraphics{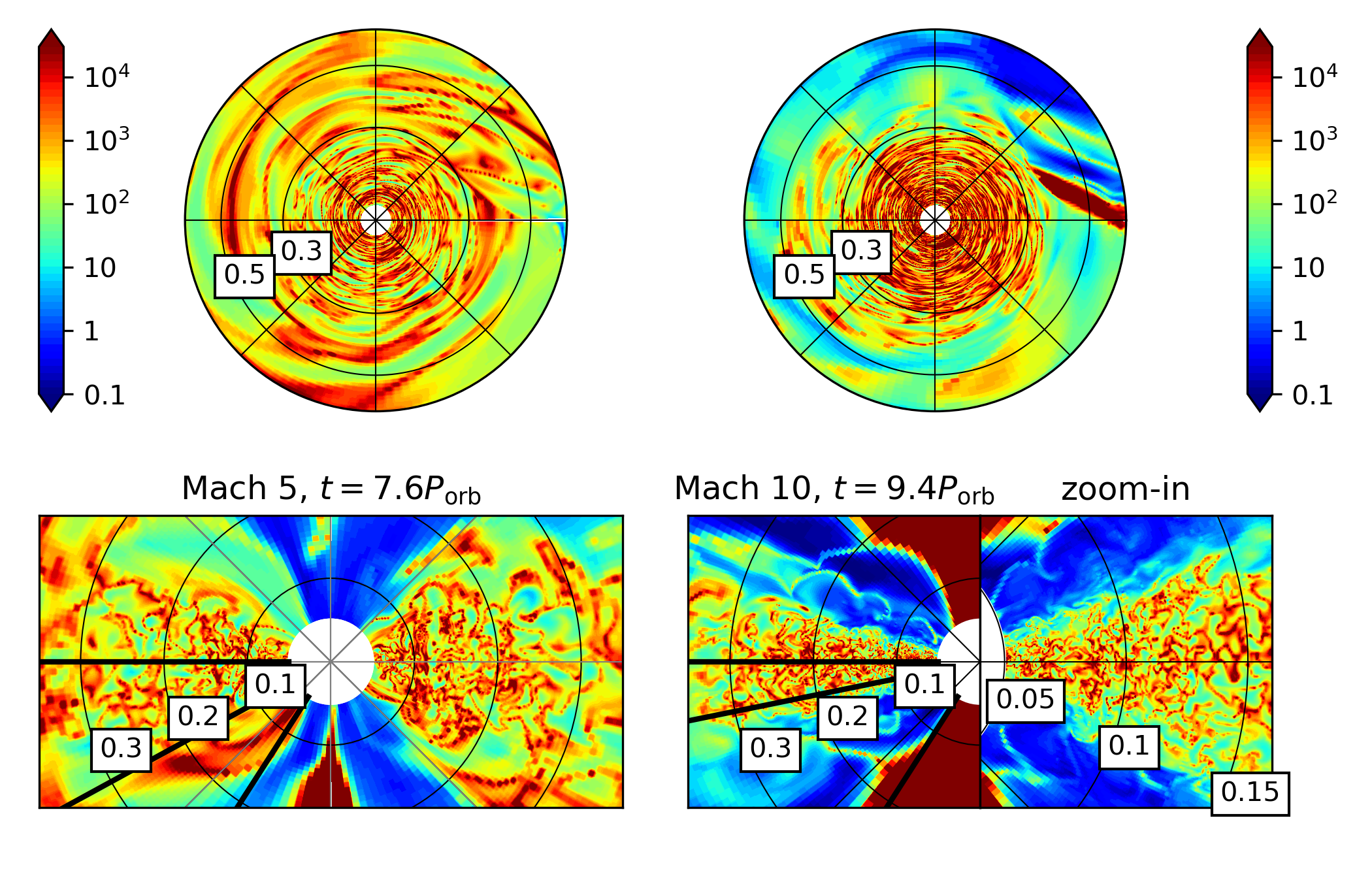}
            \caption{Plasma~$\beta$ snapshots for Mach~5 (left) and Mach~10 (right). Top: full equatorial slices, bottom: poloidal slices of the disk region ($\phi=0,180\degree$ for right and left halves, respectively). Solid black lines in the bottom plots visualize the $\theta$ ranges used to define ``disk body'' and ``corona'' regions (see Sect.~\ref{sect:results:bfield} for details).}
            \label{fig:snapshot_beta}
        \end{figure*}
    
    Fig.~\ref{fig:snapshot_beta} shows equatorial and poloidal slices of plasma~$\beta$ for snapshots at $t=7.6P_{\rm orb}$ for Mach~5 (left panels) and $t=9.4P_{\rm orb}$ for Mach~10 (right panels). A complex network of high-$\beta$ structures (current sheets) is a sign of vigorous MRI-driven turbulence. The average values of plasma~$\beta$ are consistent with zero-net-flux stratified shearing box simulations, where $\beta\sim 100-1000$ \citep[e.g.,][]{1996Stone, 2010Davis, 2016Salvesen}. In an average over the disk body ($r<0.3$, $|\theta-\pi/2| < 2.5H/R$ or $2H/R$ for Mach~5 and Mach~10, respectively) and time, we obtain $<\beta>=143$ for Mach~5 and $<\beta>=157$ for Mach~10. \cadded{We note that grid resolution of our models limits MRI dynamo operation within the disk to $\beta\lesssim1000$, above which the MRI would become unresolved.}
    
    A wealth of research has been (and continues to be) conducted on the details of MRI turbulence and the MRI dynamo, often in context of shearing box simulations (e.g., \citealt{1995Hawley, 1995Brandenburg, 2010Davis, 2010Latter}, see also reviews by \citealt{1998Balbus, 2005Brandenburg}). Global models have been studied by \cite{2011Hawley, 2012Sorathia, 2013Hawley}, with special attention given to numerical convergence. \cite{2011Hawley} proposed the following quality factors as a measure of whether the MRI is adequately resolved:
    \beq Q_i=\frac{\lambda_{\textrm{MRI},i}}{\Delta x_i} = 2\pi\sqrt{\frac{16}{15}}\frac{v_{A,i}}{\Omega_K(R)\Delta x_i}, \eeq
    where $\lambda_{\textrm{MRI},i}$; $v_{A,i}$; and $\Delta x_i$ are the most unstable MRI wavelength, Alfv\'en velocity, and cell size in the $i$-th direction of the grid, respectively, and $\Omega_K(R)$ is the local Keplerian frequency. In our models, the averages of these quality factors (over $r\in[0.1,0.25]$, $|\pi/2-\theta|\in[-H/2R,H/2R]$, and time) for Mach~5 and Mach~10 were equal to $Q_z \simeq 3.4, 3.1$ and $Q_{\phi} \simeq 12, 15$. While small, they are sufficient to describe global properties of the flow (such as plasma~$\beta$ and the $\alpha$ parameters), as shown by \cite{2012Sorathia} (see their figs.~5,8). However, the statistics of MRI turbulence cannot be resolved here. \cite{2012Sorathia} and~\cite{2013Hawley} find $H/\Delta z \gtrsim 32$, $Q_z \gtrsim 10$, and $Q_{\phi} \gtrsim 25$ to be necessary to achieve that goal (here, $H/\Delta z\simeq27$, Sect.~\ref{sect:methods:mesh}).
    
    We thus focus on the global features of magnetic field structure. As noted in Sect.~\ref{sect:results:genDescr}, the entire body of Mach~5 disk appears to be fairly well mixed (bottom left panel of Fig.~\ref{fig:snapshot_beta}), with plasma~$\beta$ at a few $100$ for most of the volume. Meanwhile, Mach~10 clearly separates into a weakly magnetized main body at $|\pi/2-\theta|\lesssim2H/R$ ($\beta\sim100-1000$) and a strongly magnetized (relative to the main body) ``corona'' at $|\pi/2-\theta|\in[2H/R,10H/R]$ ($\beta\sim$ few tens). To avoid potential confusion, we stress that the latter is \textit{not} the X-ray emitting corona as understood in observational astrophysics \citep[e.g.,][]{2013Reis, 2015Wilkins, 2016Wilkins}. Rather, this term is used as is traditional in the numerical modeling community: to denote a distinct region at the disk surface exhibiting magnetization high relative to disk body. Whether or not these two definitions are related is an interesting question in itself -- one that is, however, beyond the scope of this study. The presence of such defined ``corona'' is nearly-ubiquitous in stratified shearing box simulations since the first models presented \citep[][]{1995Brandenburg,1996Stone}. Coronal accretion has also been shown to be an important process in a number of studies, although mainly in context of non-zero net vertical flux \citep[e.g.,][]{1994Stone, 2009Beckwith, 2012Guilet, 2013Guilet, 2018Zhu}. We investigate the role of the disk ``corona'' in driving accretion in Sect.~\ref{sect:results:Mach10}.
    
    The shell-like features seen in density snapshots in Fig.~\ref{fig:snapshot_rho} are also observed in plasma~$\beta$ as strongly-magnetized regions surrounded by high-$\beta$ shells. This supports our interpretation that they correspond to weakly-magnetized gas being pushed out of the disk by buoyantly rising magnetic bubbles.
        
        \begin{figure*}
            \centering
            \includegraphics{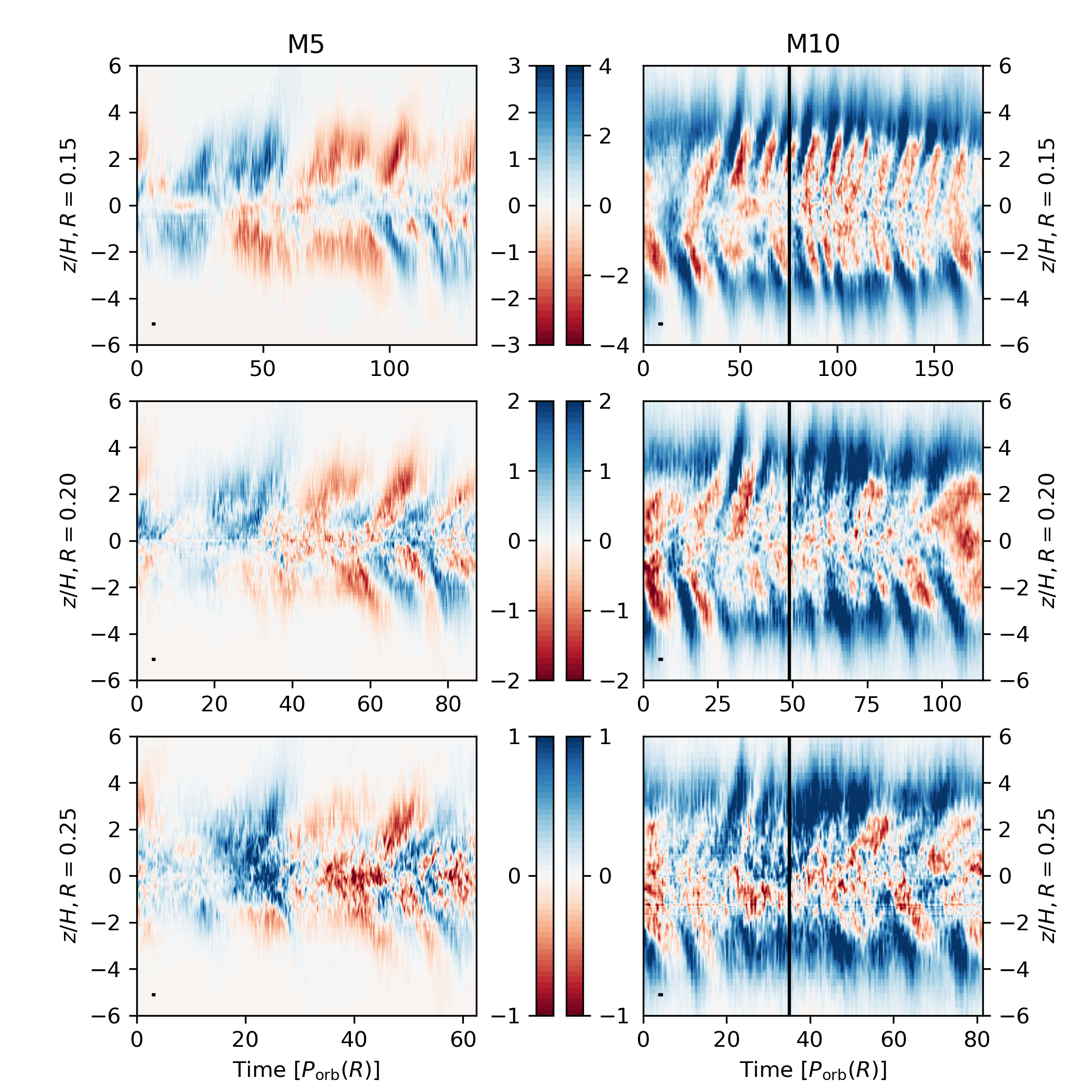}
            \caption{Butterfly diagrams for our models. Each panel shows azimuthally averaged $B_{\phi}$ as a function of height (vertical axis) and time (horizontal axis, in units of the \textit{local} orbital period). Each row corresponds to a specific radius, as indicated by captions at the left and right side of the plots. The colorbars show values of magnetic field in units of $10^{-2}$~sim.u. The top left diagram has been vertically boxcar-averaged over 3\% of its vertical extent to improve clarity. The horizontal black bar in the bottom left corner of each plot shows time resolution. \\
            \textbf{Mach~5 (left column)}: the field reversals are irregular and often asymmetric. \textbf{Mach~10 (right column)}: field reversals are fairly regular, especially at lower radii, they are not always symmetric. There is indication of change in the field reversal pattern between the two \creplaced{configuration}{phase}s of the disk.}
            \label{fig:butterfly}
        \end{figure*}
    
    In addition to magnetized coronae, another consequence of magnetic buoyancy observed in stratified shearing box simulations is the so-called ``butterfly diagram'' \citep[e.g.,][]{1995Brandenburg, 2009Shi, 2016Salvesen}. In context of magnetic field generation by the MRI dynamo, (azimuthal) magnetic field is usually produced at the disk midplane (see, however \citealt{2015Begelman}). Once the midplane field is strong enough, buoyancy and turbulent motions cause it to rise towards the surface, where Parker instability shapes a corona \citep{2009Shi}. The cycle then restarts with reversed polarity of the field. Typical duration of one such field reversal in stratified shearing box simulations is $6-10$ local orbital periods. If the vertical profile of azimuthally-averaged $B_{\phi}$ is plotted as a function of time, this results in a characteristic ``butterfly'' pattern. We plot such (butterfly) diagrams for our models in Fig.~\ref{fig:butterfly}, with each row corresponding to a different radius within the disk. \cadded{We note that, although Fig.~\ref{fig:butterfly} shows azimuthally-averaged values, the butterfly pattern is also present in our models if poloidal slices of $B_{\phi}$ are used.}
    
    For Mach~5, the field reversal patterns are irregular, often asymmetric, and at times rare. However, when they do occur, the recurrence time is consistent with $6-10$ local orbits. In previous studies, convection has been found to be able to quench field reversals in MRI turbulence \citep{2017Coleman, 2018Coleman} by transporting some of the coronal magnetic field to the midplane (other such inhibiting factors include, e.g., presence of net vertical flux, \citealt{2016Salvesen}). We speculate that large vertical displacements caused by inclined spiral shocks (see lower left panel of Fig.~\ref{fig:snapshot_rho} and Sect.~\ref{sect:results:genDescr}) may play a similar role. Turbulence itself may also contribute to this process, as the turbulent eddies are quite large in Mach~5.
    
    While the butterfly pattern for Mach~10 is more regular, here too the field reversals are often asymmetric and even missing. Highly inclined spiral shocks are still present and could be responsible. Interestingly, at $R=0.15$ (top right panel of Fig.~\ref{fig:butterfly}) the pattern changes for $t\in[75,125]P_{\rm orb}(R)$. This interval corresponds to a period of enhanced accretion (\creplaced{configuration}{phase} B) we describe in Sect.~\ref{sect:results:variability} (transition between \creplaced{configuration}{phase}s A and B at $5P_{\rm orb}$ is indicated by vertical black lines in the right panels of Fig.~\ref{fig:butterfly}). Within this time interval, the field reversals are seemingly more frequent, especially below the midplane, although they remain within the usual $6-10$ orbital periods. \cadded{The frequency of field reversals can be affected by changes in the local shear rate \citep{2015Gressel} and disk aspect ratio \citep{2018Hogg}.} Assessment of significance of \creplaced{this change}{the change we observe, however,} would \creplaced{require}{strongly benefit from} larger samples. \cadded{We thus leave a closer inspection of this behavior to a future study}.\cremoved{However, it is a potentially interesting behavior that has not previously been seen in local simulations.}
        
    \subsubsection{Spiral structure -- linear description}\label{sect:results:spiral}
  
    \begin{figure*}
        \centering
        \includegraphics{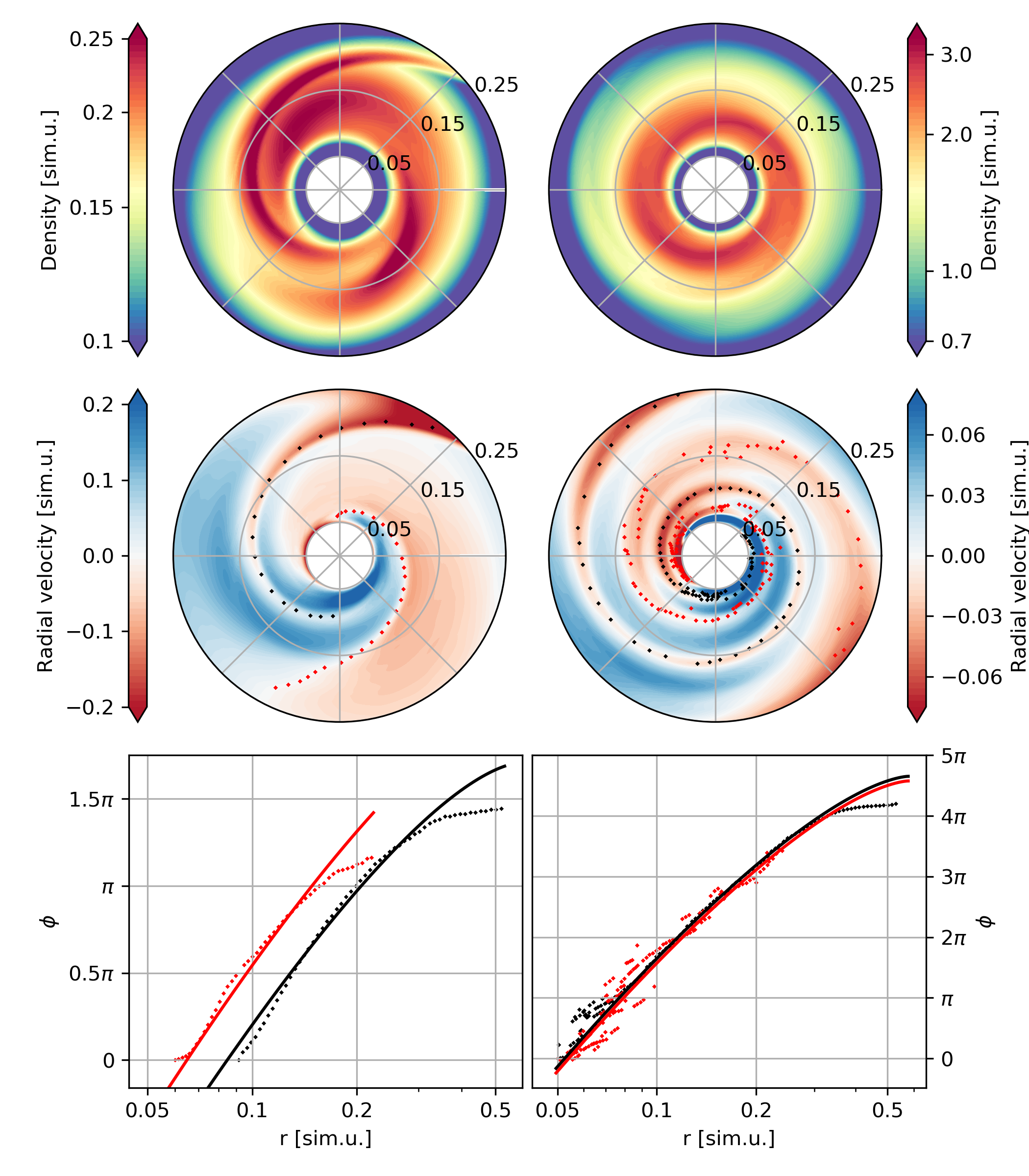}
        \caption{Time-averaged spiral structure. Left: Mach~5, right: Mach~10. Top row: time-averaged equatorial density slice. Middle row: time-averaged equatorial slice of radial velocity. Red and black diamond markers in the second and third row show local minima of the absolute value of radial velocity (``red'' arm in Mach~5) or local maxima of density (all other cases), used to track spiral arms. A limited number of points is shown for clarity. The fit of 2D spiral structure resulting from the linear dispersion relation is plotted as solid lines in the bottom row (see text for details).}
        \label{fig:spiral}
    \end{figure*}
    
    \cite{Wendy1} compare the spiral patterns in their unstratified 3D models to those expected from the linear dispersion relation for a compressible wave propagating in a 2D Keplerian flow \citep[e.g.,][]{2002Ogilvie}. They find a good fit when propagation speed is equal to the local sound speed $v_{\rm spi}=c_s$, and report further improvement for $v_{\rm spi}=\sqrt{c_s^2+v_A^2}$ (where $v_A$ is the local azimuthally-averaged Alfv\'en speed), which approximates the effect of (sub-dominant) magnetic field. It is interesting to see whether these statements hold when vertical stratification is added. Thus, Fig.~\ref{fig:spiral} focuses on the spiral structure of our models.
    
    Mach~10 clearly contains a symmetric $m=2$ tightly wound spiral. The situation in Mach~5, however, is more complex. There is a strong spiral arm associated with the inflow (top-left of the slices; black points in middle-left panel of Fig.~\ref{fig:spiral}; the ``black'' arm) and a weaker arm to the bottom-right of the plots (red points; ``red'' arm). However, we also see a spiral overdensity just to the right of the ``black'' arm (most dense at $r\sim0.1, \phi\sim0.7\pi$), wound at a different pitch angle. It is a dynamical consequence of the ``red'' arm. Each particle entering the ``red'' spiral shock at Keplerian velocity (disks orbit counter-clockwise) enters an elliptical orbit with apastron at the shock. The corresponding periastrons are located precisely at the location of the ``additional'' spiral arm, seen as the change in radial velocity sign in the middle-left panel of Fig.~\ref{fig:spiral}. Recently, \cite{2018BaeZhu} formally derived properties of such ``additional'' spiral arms in protoplanetary disks as a constructive interference between epicyclic modes excited by the main spiral arms of the system. They found these structures to merge with the primary spirals for larger (but still planetary) binary mass ratios. In our case, these ``additional arms'' are only still visible thanks to asymmetry between the ``black'' and ``red'' arms of Mach~5, caused by forced alignment of the ``black'' arm with the gas inflow. If the spiral pattern were symmetric, gas orbits would not be able to pass periastron before reaching the ``black'' arm and would merge with it as predicted by \cite{2018BaeZhu}. This highlights the benefits of using realistic feeding geometry in our models.
    
    In order to compare our spiral patterns with linear theory, we need to measure the location of these shocks. With the exception of the ``red'' arm for Mach~5 we do so by quasi-automated detection of local maxima in density, proceeding as follows.
    \begin{enumerate}
        \item The equatorial slice of density is split into separate $\rho(\phi)$ tables for each radius.
        \item The data is de-noised with the Savitzky-Golay filter \citep{1964SavGol}\footnote{As implemented in the Python SciPy library, \cite{2020SciPy-NMeth}, version 1.3.2.}. Polynomial order $3$ and window length of $\sim6\%$ of the $\phi$ range at the given radius are used.
        \item The local maxima of the resulting $\rho(\phi)$ function for each radius $r$ are saved as $(r,\phi)$ points.
        \item To aid separation of the resulting point collection into spiral arms, we use the clustering algorithm DBSCAN \citep{dbscan1, dbscan2}\footnote{As implemented in Python Scikit-Learn library, \cite{sklearn}, version 0.21.3.}. Clustering was found to be most helpful in $(\log r,\phi)$ space and $\epsilon=0.11$,~$0.15$ were set for Mach~5 and Mach~10, respectively.
        \item Finally, the resulting clusters are visually identified with spiral arms and corrected, if needed.
    \end{enumerate}
    A similar procedure is performed for the ``red'' arm of Mach~5, except that local minima of the absolute value of radial velocity have proven to be a better metric for spiral shock detection there. In that case, Savitzky-Golay smoothing is not used, and DBSCAN's $\epsilon=0.11$. We plot the resulting spiral arm locations as black and red points in Fig.~\ref{fig:spiral} (we only show every 5$^{\rm th}$ point for clarity).
    
    We then proceed to fit the shock locations with predictions from linear models. For a compressible wave with phase velocity $c_s$ propagating through a two-dimensional gas disk, the dispersion relation in the linear limit can be written as \citep[e.g.,][]{2002Ogilvie}
    \beq \left(m(\Omega-\Omega_p)\right)^2 = \kappa^2 + c_s^2k^2, \eeq
    where $\Omega$ and $\Omega_p$ denote the angular speeds of the local flow and pattern, respectively; $\kappa$ is the epicyclic frequency, $k(R)$ is the local wave number, and $R$ denotes the cylindrical radius. After \cite{Wendy1}, we use this relationship to calculate the curve of constant phase associated with a spiral wave in a nearly-Keplerian disk ($\Omega \approx \kappa$), with pattern speed set by co-rotation with the binary companion ($\Omega_p = 1$). For such a pattern, the pitch angle $\xi$ obeys \citep{Wendy1}:
    \beq\begin{array}{rl} \tan \xi(R) =& \frac{c_s(R)}{R\sqrt{(\Omega(R)-\Omega_p)^2 - \kappa^2(R)/m^2}}\nonumber\\
     =& \frac{1}{M(R)\sqrt{(1-\Omega_p/\Omega)^2-1/m^2}}, \label{eq:spiral:pitch_wendy}\end{array}\eeq
    where $M(R)$ is the local Mach number. Within our disks, sound speed is set by the \cite{ShakuraSunyaev} temperature ceiling (Sect.~\ref{sect:methods:eos}):
    \beq c_s(R) \approx c_{s,\alpha}(R) = \frac{v_K(r_{\rm in})}{M_{\rm in}} \left(\frac{R}{r_{\rm in}}\right)^{-3/8}, \label{eq:spiral:cs_locisoth}\eeq
    where $v_K(r_{\rm in})$ is the Keplerian velocity at the inner edge of the grid. At the same time, the flow is nearly Keplerian:
    \beq \kappa \approx \Omega \approx \sqrt{GM_1}/R^{3/2}, \label{eq:spiral:kappaKepl}\eeq
    \beq M(R) \approx \sqrt{\frac{GM_1}{R}} / c_{s,\alpha}(R). \label{eq:spiral:machKepl}\eeq
    Combining equations (\ref{eq:spiral:pitch_wendy})-(\ref{eq:spiral:machKepl}):
    \begin{eqnarray} \tan\xi(R) \approx \frac{\left(R/r_{\rm in}\right)^{1/8}}{M_{\rm in}\sqrt{(1-\Omega_pR^{3/2}/\sqrt{GM_1})^2-1/m^2}}, \label{eq:spiral:final}\end{eqnarray}
    which fully defines the spiral pattern $R_s(\phi_s)$ in differential form:
    \beq \frac{dR_s}{R_sd\phi_s} = \tan\xi(R_s). \label{eq:spiral_diff}\eeq
    The only free parameter is the constant of integration, i.e., azimuthal rotation of the spiral pattern as a whole $\phi_0$. To fit it to our data, we numerically integrate eq.~(\ref{eq:spiral_diff}), and minimize the quantity:
    \beq \Psi(\phi_0) = \sum_{i}D^2(P_i| S(\phi_0)), \eeq
    where summation occurs over the spiral arm points in our data $\{P_i\}$ and $D(P_i| S(\phi_0))$ is the Euclidean 2D distance between $P_i$ and its nearest point in the theoretical spiral pattern $S(\phi_0)$.
    
    Resulting fits are shown in the bottom row of Fig.~\ref{fig:spiral}. Our data are very well described by these analytical considerations, only showing deviations at the outermost points of the respective spiral arms (where our spiral arms become nearly radial). This is somewhat surprising. Our models presents a significant departure from the assumptions of eq.~(\ref{eq:spiral:final}). The flow is no longer 2-dimensional (even though we do fit to its equatorial slice), departures from Keplerian orbits can be significant (as evidenced by the ``additional'' spiral arm in Mach~5), and the pitch angles are not necessarily small. However, we welcome this finding as a validation of the use of 2D models in interpretation of physical data. Following \cite{Wendy1}, we also performed a fit with $c_s$ replaced by $\sqrt{c_s^2+v_A^2}$, approximating the effect of magnetic fields (note that $v_A << c_s$). As in their case, we do see a slight improvement of the fit. 
    
  
  \subsection{Vertical structure of the disk}\label{sect:results:vertical}
  
    \begin{figure*}
        \centering
        \includegraphics{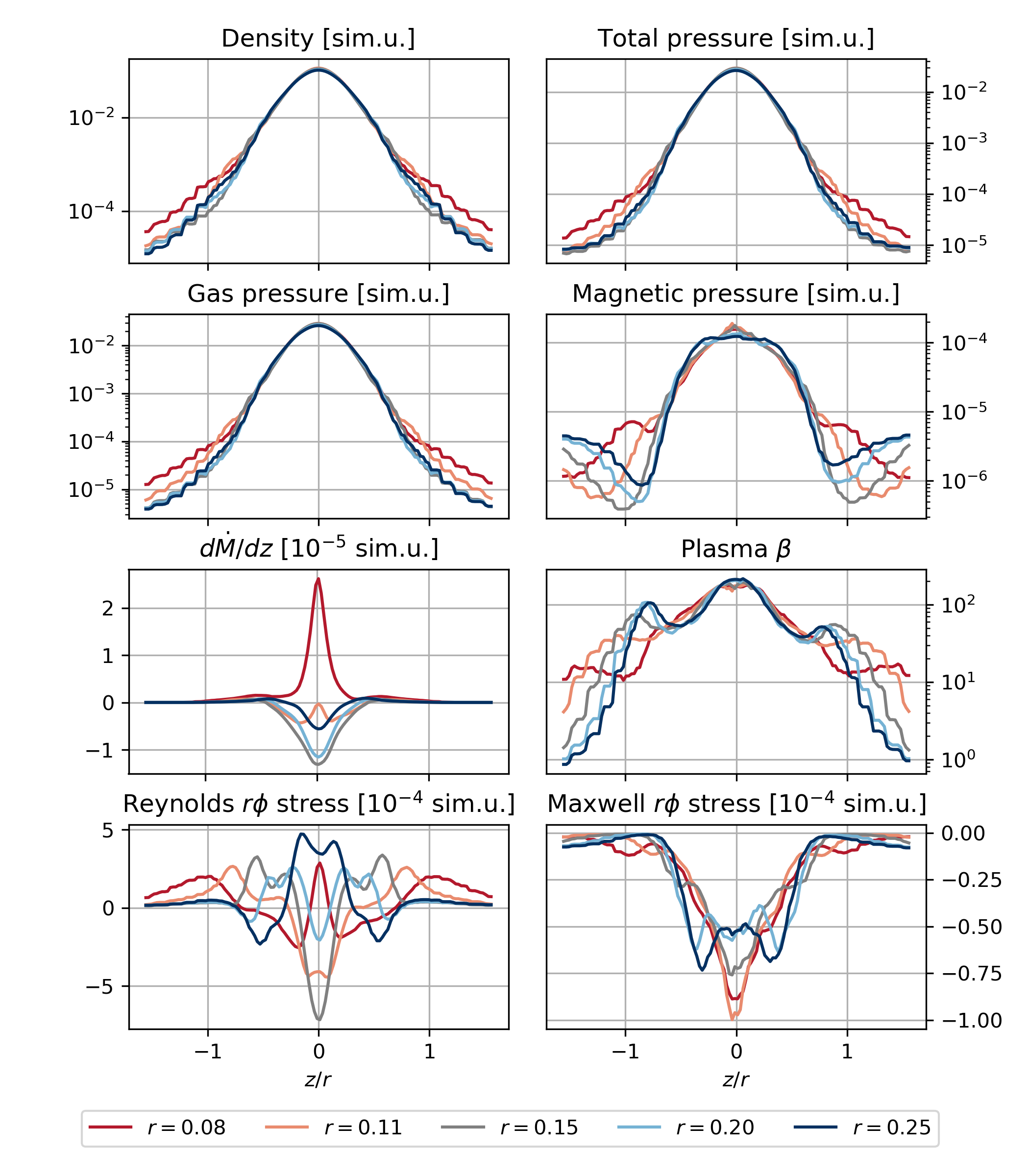}
        \caption{Time- and azimuthally-averaged vertical profiles for Mach~5. The color of each curve corresponds to the radius at which the vertical profile was measured, as indicated by the legend.}
        \label{fig:vertical_M5}
    \end{figure*}
  
    \begin{figure*}
        \centering
        \includegraphics{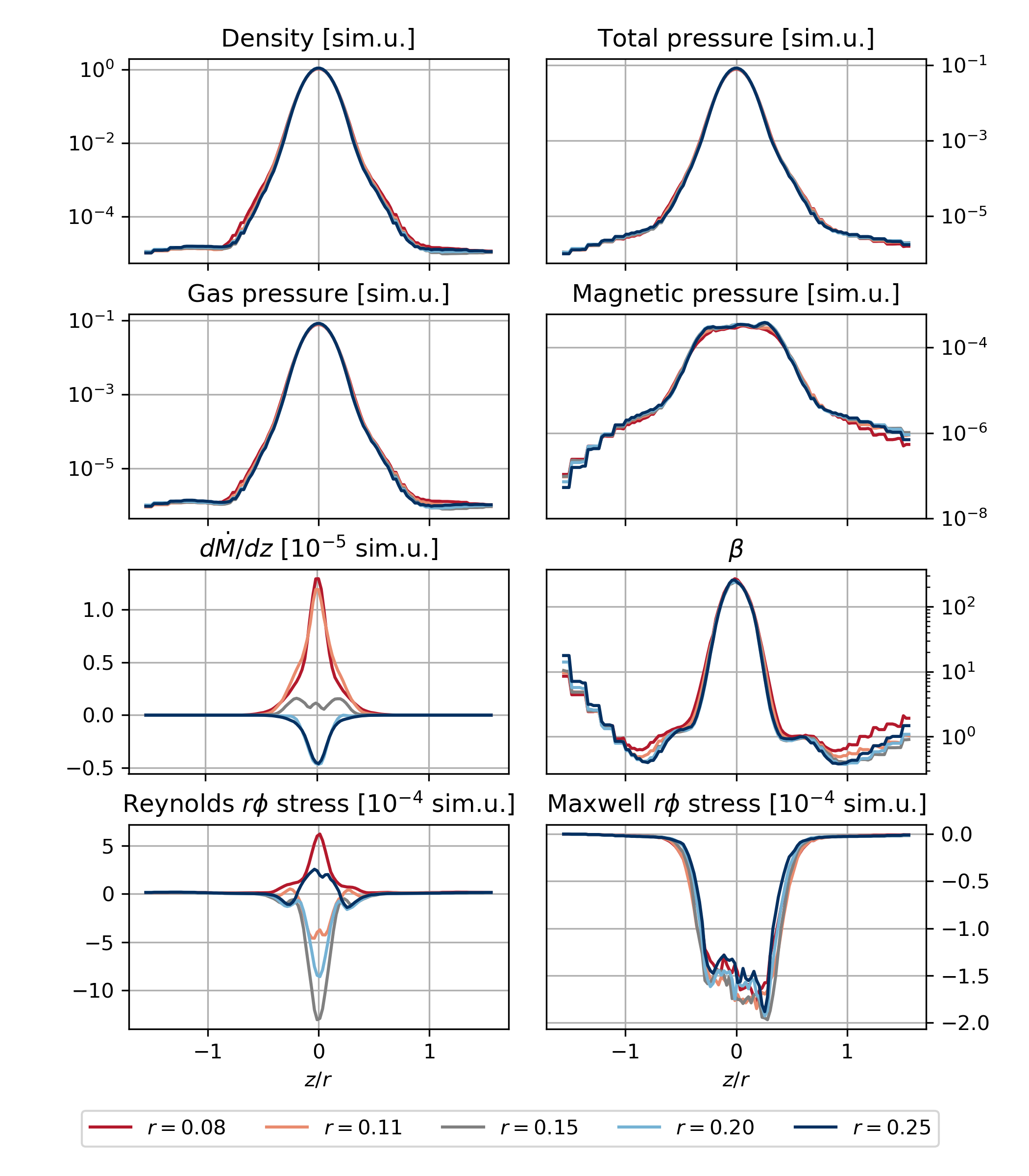}
        \caption{Time- and azimuthally-averaged vertical profiles for Mach~10. The color of each curve corresponds to the radius at which the vertical profile was measured, as indicated by the legend.}
        \label{fig:vertical_M10}
    \end{figure*}
  
  One of the main contributions of this work lies in adding vertical structure to the unstratified models of \cite{Wendy1, Wendy2}. Vertical profiles have previously been extensively studied in the local limit of stratified shearing box simulations. It is thus instructive to compare our results to these local studies.
  
  Figs.~\ref{fig:vertical_M5} and~\ref{fig:vertical_M10} show time- and azimuthally-averaged vertical profiles at several radii. For density and pressure (dominated by gas pressure), these are clearly Gaussian. This is a common finding in shearing box simulations \citep[e.g.,][]{1996Stone, 2014Hirose, 2015Begelman, 2016Salvesen} and corresponds to the assumption of vertical hydrostatic equilibrium. It is not a trivial result, as our global models could have relied on radial pressure gradients. That this is not the case aligns well with earlier global models of MRI-unstable disks in weakly-magnetized environments \citep[e.g.,][]{2017Flock, 2018Zhu, 2018Hogg, 2019Jiang}, as well as the fact that, on average, the midplane regions of our disks remain close to Keplerian rotation. Our findings thus support analytical and local numerical work, where purely vertical equilibrium is often taken as an assumption \citep[][and others]{1965Goldreich, ShakuraSunyaev, 1995Hawley, 1995Brandenburg}.
  
  Magnetic pressure profiles contain a flattened core caused by magnetic buoyancy (see Sect.~\ref{sect:results:bfield}). As a result, plasma-$\beta$ in the midplane is dominated by gas pressure with its Gaussian profile. Both these features are commonly seen in stratified shearing box simulations \citep[e.g.,][]{1996Stone, 2013Fromang, 2016Salvesen}. In the regions farthest from the disk midplane, the vertical profiles of the magnetic field differ between Mach~5 and Mach~10. For Mach~5, both $P_{\rm mag}$ and plasma~$\beta$ show variations due to long-lived turbulent structures high above the disk. We expect such fluctuations to average away in longer simulations. For Mach~10, the vertical profiles of $P_{\rm mag}$ and $\beta$ corroborate the separation into a weakly-magnetized main body and a magnetic ``corona''. In the plasma~$\beta$ profile (right plot in the third row of Fig.~\ref{fig:vertical_M10}), the weakly-magnetized main body is responsible for the central Gaussian peak, while the ``corona'' corresponds to a step-like feature at $\beta\sim1$ on either side of the midplane.
  
  Stratified box simulations sometimes report a top-hat profile for the Maxwell stress \citep{1996Stone, 2013Fromang}, which is consistent with our results for Mach~10, as well as the outer radii of Mach~5. In the latter (blue curves in the bottom-right panel of Fig.~\ref{fig:vertical_M5}), we see some evidence of two-peaked profiles, which may indicate coronal accretion. For the inner radii of Mach~5, the vertical profile of Maxwell stress becomes more peaked at the midplane.  In these inner regions, the disk scale height is not small compared to the local radius of the disk, and thus deviations from the profiles at larger radii, and in shearing box models, are not unexpected.
  
  Interestingly, we find Reynolds stress in Figs.~\ref{fig:vertical_M5},~\ref{fig:vertical_M10} to be highly variable with height and radius and often reach positive values, acting to prevent accretion in certain regions. After averaging over the disk volume, the cumulative effect of $T_R$ is thus smaller than indicated by its large peak values. Reynolds stress is also seen to act mainly in the midplane, where pressure is large, and its influence on accretion rate, measured by $\alpha_R = T_R/P$, is inhibited. In light of these two observations, although the absolute values of Reynolds stress in Figs.~\ref{fig:vertical_M5} and~\ref{fig:vertical_M10} are about a factor of 5 larger than Maxwell stress, their actual influence on disk accretion is comparable (cf. Figs.~\ref{fig:radial_M5}, \ref{fig:radialComp_M10}, and discussion in Sect.~\ref{sect:results:radial}). We further discuss the relative roles of Reynolds and Maxwell stress in driving accretion in Sections~\ref{sect:results:Mach5} and~\ref{sect:results:Mach10}. In both models, Reynolds stress is strongly driven by spiral waves. Since they are a global feature, it is not a surprise that the oscillatory vertical profiles of $T_R$ do not match the flat profiles reported by, e.g., \cite{1996Stone} in shearing box simulations of the MRI.
  
  A final point to make regarding vertical profiles is that at different radii (denoted by curves of different colors in Figs.~\ref{fig:vertical_M5} and~\ref{fig:vertical_M10}), for some variables the curves are often strikingly self-similar, as especially evident for Mach~10 (Fig.~\ref{fig:vertical_M10}). This is a strong argument confirming that vertical structure of the disk is insensitive to radius. While such independence is required by some analytical models \citep[most notably,][]{ShakuraSunyaev}, we note that it is not sufficient for their applicability. Local-only interactions, where each radial annulus of the disk interacts only with neighbouring annuli by means of viscosity, may also be necessary. As we discuss in Sect.~\ref{sect:results:radial}, this assumption is not satisfied here. In addition, the vertical profiles of $T_R$ and $\dot{M}$ are clearly not self-similar. As they are both at least partially driven by spiral structure, their behavior is specific to the global character of our models.
  
  Ultimately, treatment of radiative cooling is needed to more realistically capture vertical profiles of quantities within the disk. As radiation may transport information about disk (and white dwarf surface) conditions between distant points in the system, including it may change the picture drawn by the results presented here. \cadded{Realistic disk thermodynamics would also allow convection to occur. If present, it can significantly enhance the Maxwell stress-to-pressure ratio $\alpha_M$ at certain disk temperatures \citep[e.g.,][]{2016Coleman, 2018Coleman, 2018Scepi_conv}.} We are looking forward to investigating these effects as part of our future work (see Sect.~\ref{sect:conclusions}).
  
  \subsection{Radial structure of the disk}\label{sect:results:radial}
    
    \begin{figure}
        \centering
        \includegraphics{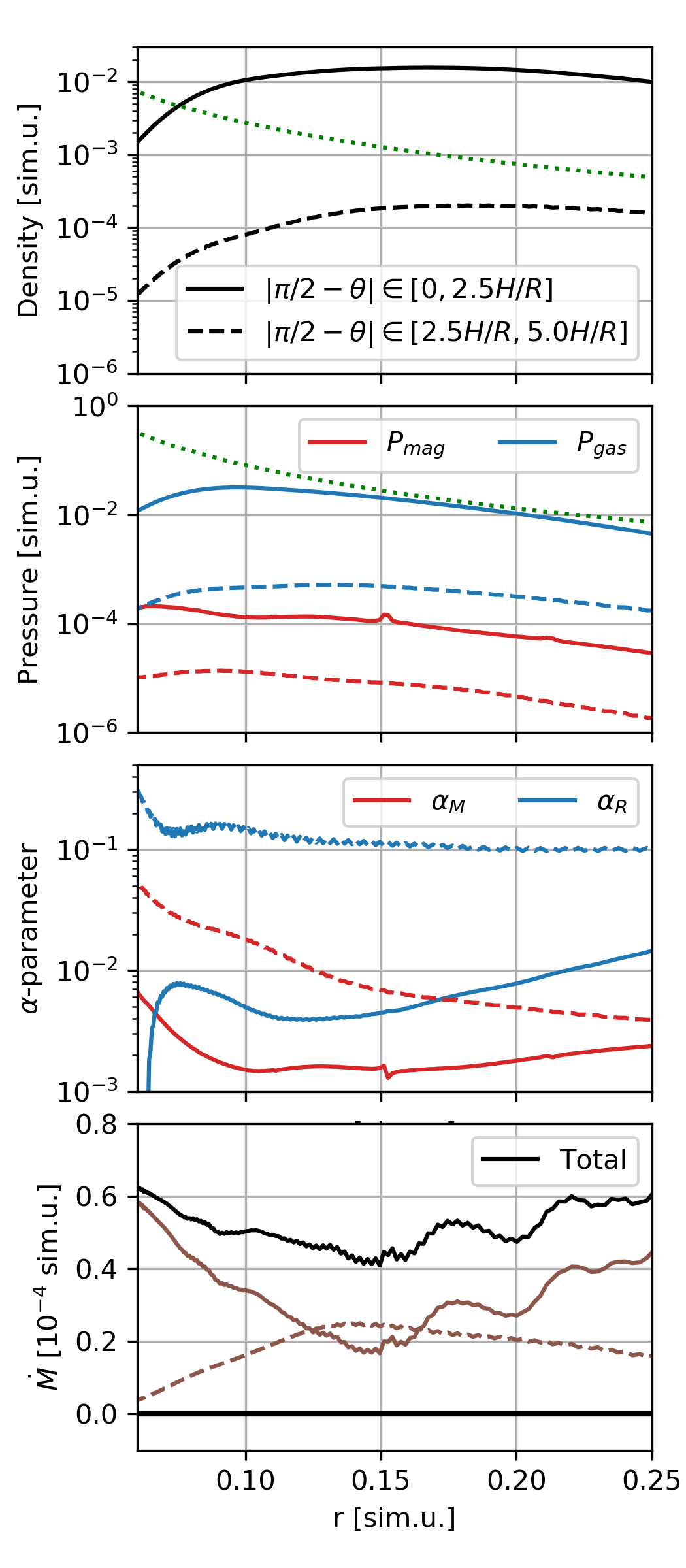}
        \caption{Time-averaged radial profiles for Mach~5. Solid and dashed curves correspond to the disk's ``main body'' and ``corona'', respectively. \cite{ShakuraSunyaev} slopes for density ($\rho\propto r^{-15/8}$) and pressure ($P\propto r^{-21/8}$) are shown as green dotted curves. Accretion is driven by Reynolds stress at all radii, with midplane-accretion dominating the innermost regions of the disk.}
        \label{fig:radial_M5}
    \end{figure}
  
    \begin{figure*}
        \centering
        \includegraphics{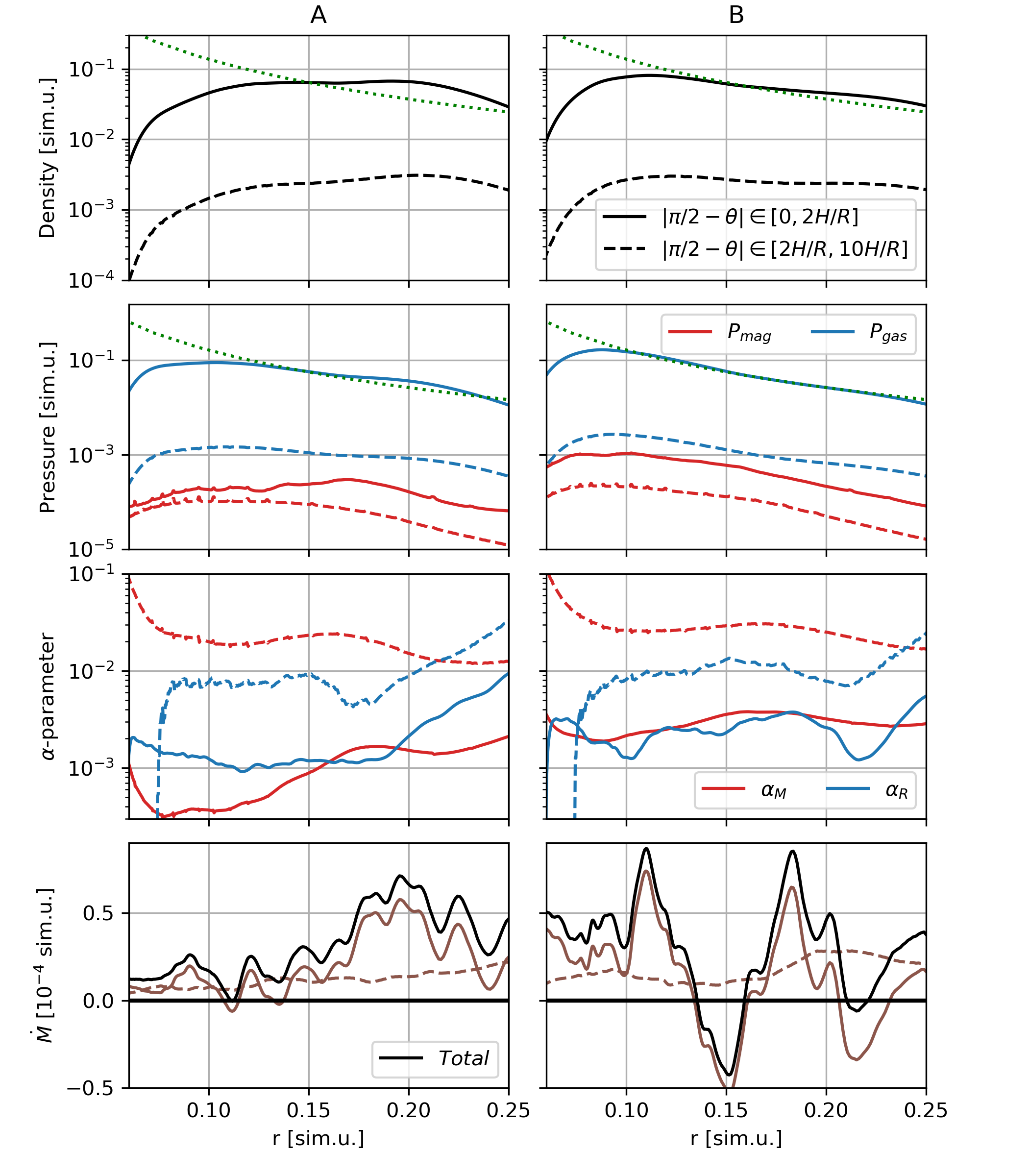}
        \caption{Time-averaged radial profiles for Mach~10 at disk \creplaced{configuration}{phase}~A (left) and~B (right). Solid and dashed curves correspond to the disk's ``main body'' and ``corona'', respectively. \cite{ShakuraSunyaev} slopes for density ($\rho\propto r^{-15/8}$) and pressure ($P\propto r^{-21/8}$) are shown as green dotted curves. The long-time ($\sim 10P_{\rm orb}$) variability causing the transition between the two \creplaced{configuration}{phase}s is clearly associated with increase in magnetic field strength in \creplaced{configuration}{phase}~B.}
        \label{fig:radialComp_M10}
    \end{figure*}
  
  As discussed in Sect.~\ref{sect:methods:eos}, our temperature ceiling corresponds to a gas pressure dominated $\alpha$-disk with free-free opacity. For such a disk, the $\alpha$-prescription predicts density and pressure to follow $\rho\propto r^{-15/8}$ and $P\propto r^{-21/8}$. These slopes are plotted (with arbitrary normalization) as green dotted lines in Figs.~\ref{fig:radial_M5} and~\ref{fig:radialComp_M10}. Generally, the \cite{ShakuraSunyaev} model describes the gas (total) pressure profiles of the disks main-body (solid lines in Figs.~\ref{fig:radial_M5} and~\ref{fig:radialComp_M10}) fairly well for $r\gtrsim 0.12$. However, only the main-body density profile for \creplaced{configuration}{phase}~B of Mach~10 (top right panel of Fig.~\ref{fig:radialComp_M10}, solid curve) and the outer radii of the Mach~5 disk (top panel of Fig.~\ref{fig:radial_M5}, solid curve) are fit well. The different slope of $\rho$ in \creplaced{configuration}{phase}~A of Mach~10 likely results from the difference in accretion rate between inner and outer radii of the disk (see Sect.~\ref{sect:results:variability}). At $r\lesssim0.12$, both pressure and density drop inwards for all our models. At that point, the length scales of the problem approach the local radius, which breaks the $\alpha$-model's locality assumptions. The same applies to coronal quantities (dashed curves in Figs.~\ref{fig:radial_M5} and~\ref{fig:radialComp_M10}), where large vertical extent of the structures (comparable with $R$) is inconsistent with the assumptions of \cite{ShakuraSunyaev}. These differences are not unexpected, and we stress that the $\alpha$-disk model remains a good description of the radial structure of our flow in regions where its assumptions are satisfied.  This is in line with other studies where an agreement with \cite{ShakuraSunyaev} is reported in appropriate regimes. This is seen, for instance, in the limit of weak wind-driving in the models of \cite{2019Scepi} (see their fig.~2) or the outer regions of semi-global simulations of \cite{2018Hogg}. We note that the level of alignment with the \cite{ShakuraSunyaev} models may change with more realistic treatment of the system. On one hand, radiative cooling (planned for our future work, see Sect.~\ref{sect:conclusions}) will introduce non-local interactions that may break the locality assumption of $\alpha$-models. On the other, accretion disks in astrophysical semi-detached binaries are colder and geometrically thinner than those presented here (as discussed in Sect.~\ref{sect:intro:limitations}), which alleviates some of the tension related to large $H/R$ ratio, bringing true accretion disks closer to the \cite{ShakuraSunyaev} solution.
  
  \subsubsection{Spiral structure drives accretion at very low Mach numbers}\label{sect:results:Mach5}
  
  As we discussed in Sect.~\ref{sect:intro:limitations}, numerical resources limit our global models to accretion disks that are hotter (of $M_{\rm in}=5,10$) than real astrophysical systems (e.g., CVs, where $M_{\rm in}\gtrsim50$). While the Mach~5 model is unlikely to be realized in nature, it can provide a useful context to the Mach~10 data.
  
  With this in mind, we use Figs.~\ref{fig:vertical_M5} and~\ref{fig:radial_M5} to explain the mechanisms driving accretion in our hot disk model Mach~5. As seen in Fig.~\ref{fig:radial_M5}, the body of the disk (solid brown line, $\Theta\lesssim 2.5H/R$) dominates the accretion rate for most of the grid. Its $\dot{M}$ has a minimum at $r\sim0.15$, where its impact becomes equal to that of the ``corona'' (dashed brown line, $\Theta\in[2.5,5]H/R$), the latter's $\dot{M}$ being maximal at $r\sim0.15$. This radius also corresponds to the transition from two-peaked to triangle-like vertical profile in Maxwell stress (Fig.~\ref{fig:vertical_M5}, see Sect.~\ref{sect:results:vertical}) -- a likely related observation. In Fig.~\ref{fig:radial_M5} we see that Reynolds stress dominates over Maxwell stress both within the main body and the ``corona''. In the former, the difference is about a factor of $2-3$, comparable with the MHD model of \cite{Wendy1}. The radial velocity slice in Fig.~\ref{fig:spiral} associates this Reynolds-stress-driven accretion with grand-design spiral structure, pointing to it as the main accretion driver in Mach~5.
  
  \subsection{The long-term evolution of Mach~10 model}\label{sect:results:Mach10}
    
    \begin{figure}
        \centering
        \includegraphics{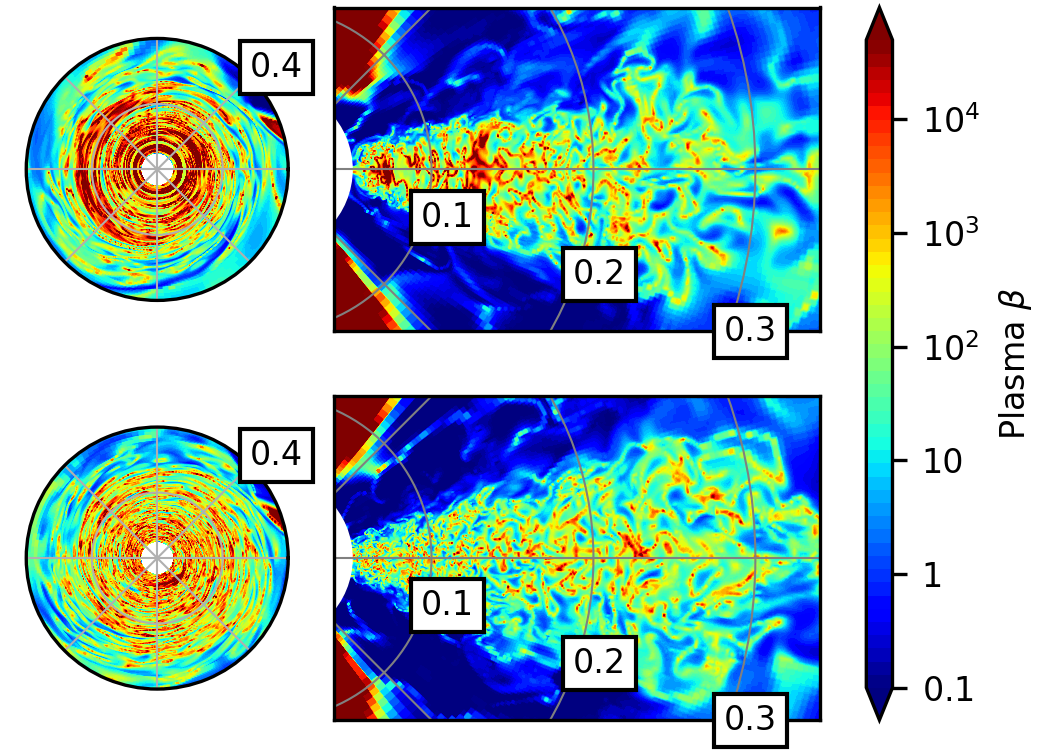}
        \caption{Plasma~$\beta$ snapshots for Mach~10, \creplaced{configuration}{phase}~A (top) and~B (bottom). Left: equatorial slice, right: poloidal slice ($\phi=180\degree$). The strengthening of main-body magnetic field in \creplaced{configuration}{phase} B is clearly visible. Note that the flow is MRI-turbulent in both cases.}
        \label{fig:betaComp_M10}
    \end{figure}
    
    \begin{figure*}
        \centering
        \includegraphics{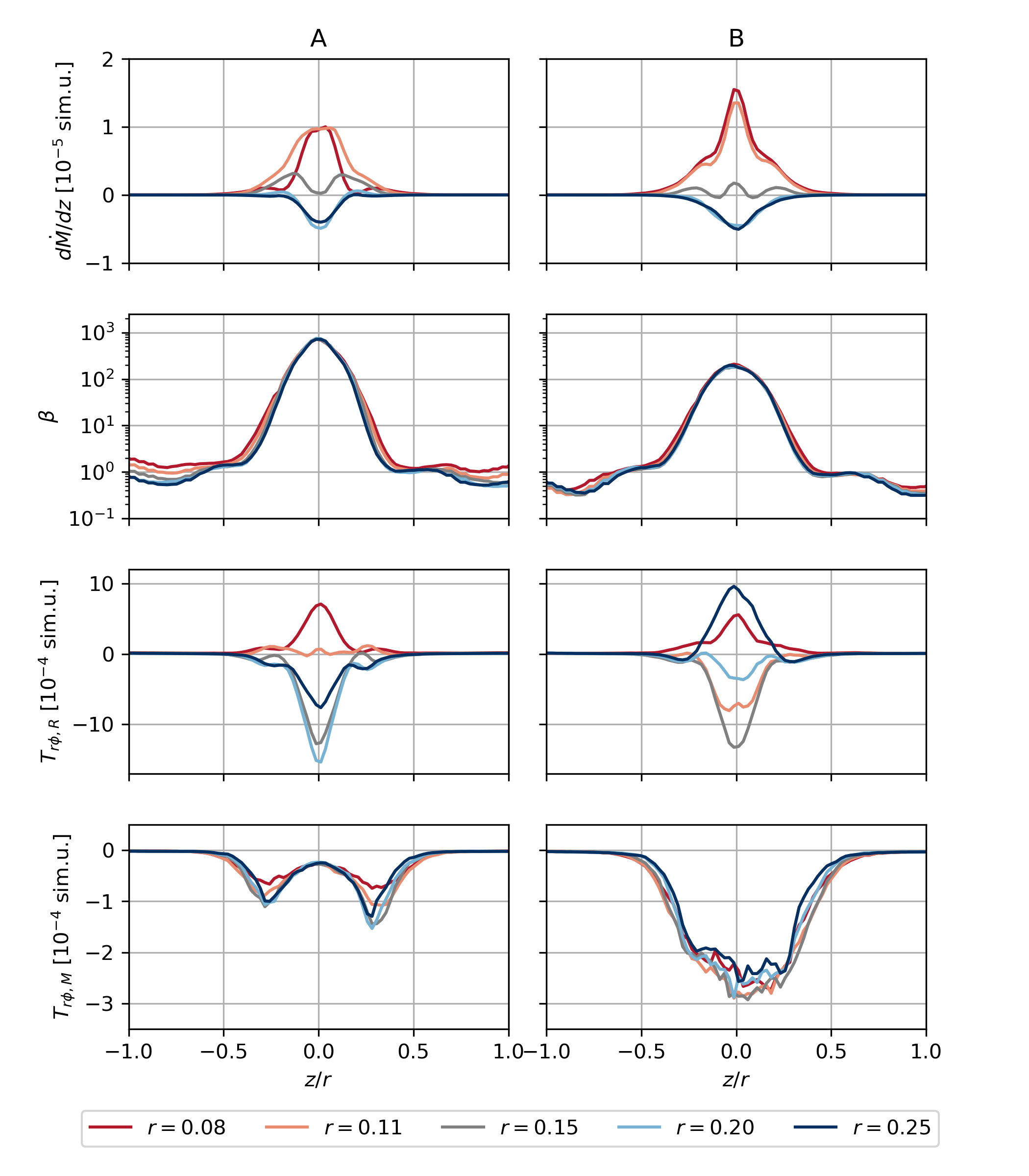}
        \caption{Time- and azimuthally-averaged vertical profiles for the Mach~10 run, where \creplaced{configuration}{phase}s A (left) and B (right) are treated separately. The color of each curve corresponds to the radius at which the vertical profile was measured, as indicated by the legend.}
        \label{fig:verticalComp_M10}
    \end{figure*}
  
  \creplaced{While Mach~10 does not reach a steady-state}{Within our current computational limits on Mach~10 simulation's runtime, the disk is not seen to reach a statistcally steady state. However, as we are only able to study a single viscous time scale of the disk (see Sect.~\ref{sect:results:variability}), it is entirely possible that such a state would be achieved if the model was allowed to continue. Still}, the long timescale ($\sim10P_{\rm orb}$) variability observed in \creplaced{this}{our Mach~10} model may be of interest for understanding disk accretion. Not only is it potentially present during a change in disk accretion rate, but there is some indication that it may be recurring in nature (see Sect.~\ref{sect:results:variability}). In this section, we attempt to understand the differences between \creplaced{configuration}{phase}s A and B (Fig.~\ref{fig:mdot}).
  
  Let us first establish whether the changes are driven by the main body of the disk or by its ``corona''. Fig.~\ref{fig:radialComp_M10} shows the radial profiles of the Mach~10 run, time-averaged separately for \creplaced{configuration}{phase}s A and B. The coronal accretion rate (bottom row, dashed brown curve) seems to remain unchanged at a level of $\sim2\times10^{-5}$. It is driven by Maxwell stress (third row of Fig.~\ref{fig:radialComp_M10}), with $\alpha_M$ $2-3$ times larger than $\alpha_R$ for $r\lesssim0.2$. As the coronal $\alpha_M$ and $\alpha_R$ are similar for \creplaced{configuration}{phase}s~A and~B, it must be the main body of the disk that drives changes in accretion rate. In \creplaced{configuration}{phase} A, $\alpha_R$ remains nearly constant with radius at $\sim1\times10^{-3}$, while $\alpha_M$ is very small close to the inner disk edge, slowly growing with radius up to $r\sim0.17$, where it flattens at $\sim2\times10^{-3}$. Meanwhile, in \creplaced{configuration}{phase} B, both main-disk $\alpha$ parameters are nearly equal throughout the disk at $\sim2-3\times10^{-3}$. We note that this latter situation is very reminiscent of the results of \cite{Wendy2}, who found Reynolds and Maxwell stress to play comparable role in driving accretion in their models at $M_{\rm in} = 10$ (note, however, that even in \creplaced{configuration}{phase} B, the coronal accretion, absent in their models, still drives $\sim1/3$ of our total accretion rate). The enhancement in $\alpha_M$ is accompanied by a factor of two increase in the main-disk $\dot{M}$ in \creplaced{configuration}{phase}~B, causing a $50\%$ rise in the total accretion rate. The main disk magnetic field is also stronger in \creplaced{configuration}{phase}~B, as shown in Fig.~\ref{fig:betaComp_M10}.
  
  In Fig.~\ref{fig:verticalComp_M10}, the Maxwell stress vertical profile is seen to change from two-peaked (efficient coronal accretion) for \creplaced{configuration}{phase}~A to a top-hat profile (main-body accretion) in \creplaced{configuration}{phase}~B (familiar from stratified shearing box simulations, e.g.,  \citealt{1996Stone}). Reynolds stress also increases, pointing to the interplay between MRI turbulence and spiral shocks reported by \cite{Wendy1, Wendy2}.
  
  The transition from \creplaced{configuration}{phase} A to B thus appears to be driven by increased MRI activity in the disk body (Fig.~\ref{fig:betaComp_M10}), which also acts to enhance the role of spiral shocks \citep{Wendy1, Wendy2} -- both of which conspire to raise accretion rate in the disk midplane. To our knowledge, this is the first time such an event is observed in a global MRI-unstable model of an accretion disk and it is likely to be related to our continuous supply of mass through the Roche-lobe overflow (which distinguishes our models from previous MHD work). The exact underlying cause for this behavior remains elusive to us. The duration of this transition is much longer than MRI saturation times within the disk (which should be limited to at most $\sim15$ \textit{local} Keplerian orbits) and MRI turbulence is well established in both \creplaced{configuration}{phase}s A and B (see Fig.~\ref{fig:betaComp_M10}). It is thus unlikely that the transition is related to any form of MRI turbulence growth. Relation to magnetic field growth through the MRI dynamo also appears unlikely, as many field reversal cycles are seen in Fig.~\ref{fig:butterfly} during \creplaced{configuration}{phase}~B of Mach~10. At face value, it appears that a ``preferred'' accretion rate exists within the disk (realized in \creplaced{configuration}{phase} A, see Fig.~\ref{fig:mdot}) and it is only surpassed once density and magnetic field accumulate beyond a certain level. In our models, this first happens at the outer edge and consequently starts \creplaced{a snowball effect}{an avalanche} as the additional mass accretes through smaller radii, removing accumulated mass (at an accretion rate higher than that of the inflow), restoring initial density levels, and allowing for resumption of accretion at the ``preferred'' level. Such disk-specific accretion rate could perhaps be set by magnetization of the inflow and / or the disk itself, the latter in turn depending on the disk Mach number and interactions with the magnetized ``corona''. \cadded{More realistic cooling and the resulting convection could enhance MRI-driven accretion rate \citep{2018Scepi_conv, 2018Coleman} increasing such a ``preferred'' level of accretion. Thus, if the above explanation is correct, more realistic thermodynamics would either decrease the difference in $\dot{M}$ between phases A and B or cause their duration to be longer.}
  
  If present in real semi-detached binaries, such an outside-in mechanism for ``waves'' of enhanced midplane accretion rate could perhaps work alongside thermal instability of the disk, assisting in propagation of an ionization front during a state change in dwarf novae. Future studies, with longer run times and higher resolution, are needed to assess what role such episodes may play in true accretion disks and potential steady states of semi-detached binaries.
  
  \subsection{Observational appearance}\label{sect:results:obs}
  
  Our models exhibit strong variability at a wide range of time scales (Sect.~\ref{sect:results:variability}). However, it is difficult to assess how the changes in local accretion rates (Fig.~\ref{fig:mdot}) translate into variability of the observed disk-integrated spectrum. In optically thick disks, it is not obvious how the increase in accretion rate or midplane density would modify the disk appearance (as set at the photosphere). Most likely, the local temperature and position of the disk photosphere is modified, resulting in observational changes. As our present models feature no radiative transport or realistic thermodynamics, we thus leave variability assessment to future studies (see Sect.~\ref{sect:conclusions}).
  
  \begin{figure}
      \centering
      \includegraphics{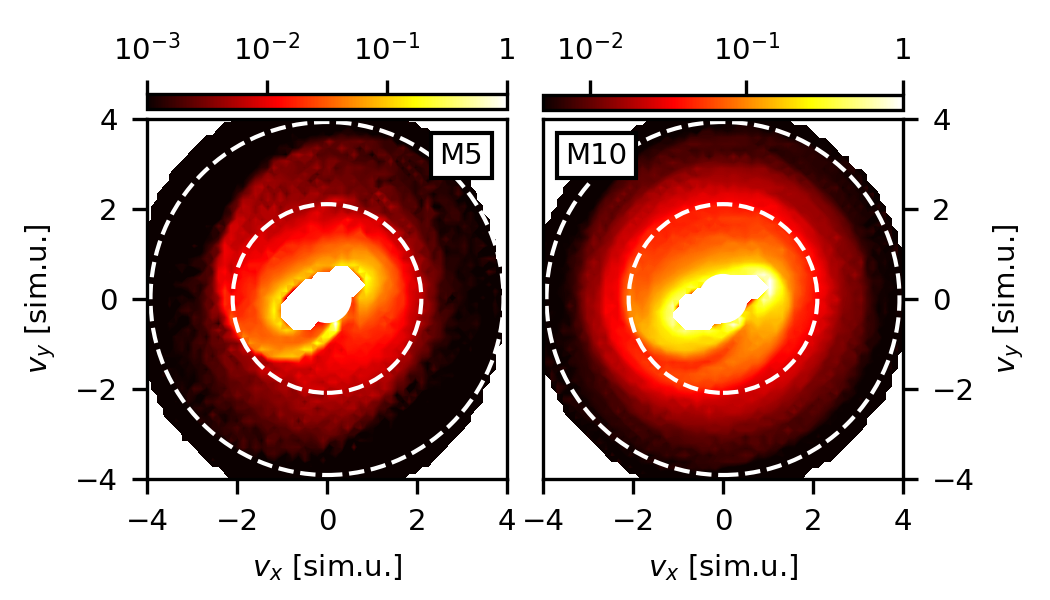}
      \caption{Edge-on Doppler diagrams for Mach~5 (left) and Mach~10 (right) constructed from our data (see text for details). The color scale is logarithmic in arbitrary flux units, as shown by the colorbars. The two dashed-line white circles in each panel correspond to Keplerian velocities at the $r=0.05$~sim.u. (inner edge of the disk, outer circle) and $r=0.3$~sim.u. (outer edge of the disk, inner circle), defining the approximate annulus of data from the accretion disk. The inflow and spiral structure are both clearly visible in both plots.}
      \label{fig:doppler}
  \end{figure}
  
  Spiral structure has been extensively documented in CVs by means of eclipse mapping and Doppler tomography \citep[e.g.,][]{1997Steeghs, 2005Baptista, 2006Klingler, 2008Khruzina, 2020RuizCarmona}. It is then of interest to ask how the spiral structure of our models of semi-detached binaries (Sect.~\ref{sect:results:appearance},~\ref{sect:results:spiral}) would be seen in such observations. Due to our approximate treatment of temperature (Sect.~\ref{sect:methods:eos},~\ref{sect:results:genDescr}), we cannot represent any non-axisymmetric distribution of surface brightness on our disks. However, we can quantify the influence of the photosphere's shape and gas velocity on observations. As a proof of concept, we attempt to generate a Doppler diagram from our data. We use isocontours of time-averaged density at $\rho=0.015$ as an approximate handle on the location of the photosphere. For each isosourface point $P_{i,j}$, we approximate the corresponding area element $\Delta S_{i,j}$:
  \beq\begin{array}{rl}
  \Delta S_{i,j} \simeq & 0.5\left(l(P_{i,j},P_{i-1,j}) + l(P_{i,j},P_{i+1,j})\right) \\
  \times & 0.5\left(l(P_{i,j},P_{i,j-1}) + l(P_{i,j},P_{i,j+1})\right),
  \end{array}\eeq
  where $l(P_{i,j},P_{k,l})$ is the 3D distance between points $P_{i,j},P_{k,l}$, and close indices correspond to neighbouring points. We assume the observed surface brightness to be proportional to the local $T^4\propto c_s^2$ (where we use time-averaged sound speed $c_s$), so the flux associated with each point is proportional to $dS\times c_s^2$. We then calculate cartesian velocities of the fluid $v_x$ and $v_y$ for each point on the isocontour (using time-averaged velocities from our simulations), and bin them into $64$ velocity channels for each direction. Finally, we sum the total ``flux'' ($dS\times c_s^2$) corresponding to each such bin in velocity space, obtaining a $64\times64$ Doppler diagram of the disk seen nearly edge-on.
  
  The resulting Doppler diagrams are shown in Fig.~\ref{fig:doppler}. Outer edges of the disk appear bright at the center of these velocity-space plots due to their large emitting surface. The inflow is well visible as a bright line near the diagrams' centers, starting slightly below $(v_x,v_y)=(0,0)$ and moving down and to the left. The disk's Keplerian velocities correspond to the space between two dashed white circles in the diagrams in Fig.~\ref{fig:doppler}. Once the inflow reaches these regions, it is seen to connect to the spiral structure, which is also clearly recognizable in the plots. We thus confirm that, at least under the assumptions of this extremely simple observation model, the structure of the disk we describe in Sect.~\ref{sect:results:appearance} could indeed be potentially observable via Doppler tomography. We note that in real semi-detached binaries disk Mach numbers are higher than in our models (Sect.~\ref{sect:intro:limitations}) and thus the spiral structure is more tightly wound than can be seen in Fig.~\ref{fig:doppler}. Moreover, outer edge of the disk (the innermost regions of Doppler diagrams) is generally too dim to be observed in as much detail as can be seen here, and the diagrams themselves would generally be available in lower resolution than allowed by our models. With these caveats in mind, however, our Fig.~\ref{fig:doppler} can be indeed found similar to the Doppler diagrams of, e.g., V2051~Oph by \cite{2016Rutkowski} (see their fig.~7) or EC21178-5417 by \cite{2020RuizCarmona} and \cite{2020Khangale} (fig.~6a). We also note that our artificial Doppler diagrams are similar to those obtained with hydrodynamical models by, e.g., \cite{1999Matsuda, 1999Haraguchi, 1999Steeghs, 2001Kunze, 2003Lanzafame, 2004Foulkes}. This supports our discussion on similarities of the general flow properties with hydrodynamical models in Sect.~\ref{sect:results:genDescr}.

\section{Conclusions}\label{sect:conclusions}

    \cadded{While spiral shocks can provide angular momentum transport in accretion disks at very low Mach numbers ($M_{\rm in}\lesssim 10$), the MRI is likely necessary to drive accretion in colder fully-ionized environments \citep[e.g., cataclysmic variables,][]{Wendy2}, potentially in concert with other mechanisms, such as magnetic interactions with disk winds. Thus, there is a need to augment the large body of work concerning global hydrodynamical models of semi-detached binaries with MHD, to include this transport mechanism self-consistently.}

    \cadded{To address this need, }we have performed the first stratified global MHD simulations of accretion disks fed by an accretion stream due to Roche lobe overflow. In doing so, we can for the first time observe the MRI turbulence self-consistently interact with fully three-dimensional global accretion disk structure in idealized models of semi-detached binaries. Despite limitations inherent to large numerical studies (which we discuss in Sect.~\ref{sect:intro:limitations}), we find robust global behaviors in our idealized models that may be helpful in understanding accretion in true semi-detached binaries, such as Cataclysmic Variables. Thus, focusing on global dynamics, we report several interesting observations from our results for models with Mach numbers of~5 and~10:
    \begin{enumerate}
        \item Accretion rate through both disks is found to be extremely variable at all time scales. While Mach~5 reaches a quasi-stationary state, Mach~10 exhibits $\dot{M}$ variability even at the time scale of the entire simulation window we analyze ($\sim10$ binary orbits).
        \item Both disks exhibit spiral structure, with position and inclination of spiral shocks changing rapidly. The midplane slices of time-averaged spiral structure, however, are fit extremely well with two-dimensional linear dispersion relation for a compressible (sound) wave propagating through a Keplerian disk \citep[e.g.,][]{2002Ogilvie}.
        \item Mach~10 is clearly separated into a gas-dominated disk body below $\sim2$ thermal scale heights and a strongly-magnetized (relative to the disk main body) ``corona''. Coronal accretion is seen to provide $30-50\%$ of the total accretion rate.
        \item The butterfly diagrams of our models are fairly irregular, asymmetric, with field reversals occasionally absent. We hypothesise that the absence of some field reversals may be related to inclined spiral shocks mixing the disk and coronal magnetic field, similar to convective motions in \cite{2017Coleman}.
        \item Our results show many similarities to stratified shearing box models. \cadded{The vertical profiles of density, gas, and total pressure are gaussian, while Maxwell stress follows a top-hat vertical profile at most radii. The disk separates into a weakly-magnetized main body and a magnetic ``corona'', with the butterfly diagram at times showing field reversals every 6-10 local orbital periods.} \creplaced{although}{However,} some effects (e.g., those attributed to spiral structure) are clearly global. \cadded{The Reynolds stress is highly oscillatory with radius and height, locally reaching positive values. Close to the disk's inner edge and in the corona, we see densities and pressures lower than those from $\alpha$-prescription, and the Maxwell stress' vertical profile can become triangular in shape.}
        \item The longest time scale ($\sim10P_{\rm orb}$) variability in Mach~10 is an outside-in accretion event (akin to \creplaced{a ``snowball effect''}{an avalanche}) seen as a temporary enhancement in midplane accretion through MRI turbulence (which in turn enhances spiral shock accretion, see \citealt{Wendy1, Wendy2}). We see some indication of a recurrent nature of these events. Longer, higher-resolution studies are needed to verify their role in true systems, and potential existence of steady states at high Mach numbers.
    \end{enumerate}
    Finally, we attempt to relate our simulations to observational results, constructing a toy-model observation of a Doppler diagram from our data. Despite simplicity of this procedure and our use of Mach numbers lower than expected in accretion disks realised in nature, we reach some qualitative agreement with the main features observed in Doppler diagrams of CV disks \citep[e.g.,][]{2016Rutkowski, 2020RuizCarmona, 2020Khangale}. \cadded{While we use CV observations for comparison, we note that our models are not limited to WD primaries and can in principle be used (with their limitations kept in mind, Sect.~\ref{sect:intro:limitations}) to describe accretion disks in other semi-detached binaries.}
      
    Our current models feature a very simple treatment of disk temperature, where an adiabatic equation of state is limited by radius-dependent temperature floor an ceiling. In future work, we intend to replace this prescription with realistic radiative cooling using the radiative transfer module of \texttt{Athena++}. In addition to our ability to track disk temperature self-consistently, this extension will enable us to construct mock observations from our data, providing a potent means of comparison with real astrophysical objects.


\acknowledgments

We would like to thank Jenny Greene, Matthew Kunz, Anatoly Spitkovsky, Omer Blaes, and Charles Gammie, for helpful discussions throughout the duration of this project. We also thank the anonymous reviewer for their many helpful comments, which have greatly improved this manuscript.

This research has been supported by NSF grant AST-1715277 ``Collaborative Research: Predicting the Observational Signatures of Accreting Black Holes'' awarded to J.~M.~Stone, and by Princeton University.

The simulations presented in this work were partly performed on computer systems generously provided by the Princeton Institute for Computational Science and Engineering (PICSciE). The authors also gratefully acknowledge the HPC resources (Stampede2) provided by the Texas Advanced Computing Center (TACC) at The University of Texas at Austin (\url{www.tacc.utexas.edu}), awarded through the Extreme Science and Engineering Discovery Environment (XSEDE) grant TG-AST190054. XSEDE \citep{xsede} is supported by NSF grant ACI-1548562.

The 3D rendering in Fig.~\ref{fig:3D} of this paper has been generated using Mayavi \citep{mayavi}.


\added{\software{\texttt{Athena++} \citep[modified development version, see Sect.~\ref{sect:methods};][]{2020StoneAthenaPP},
SciPy \citep[version 1.3.2,][]{2020SciPy-NMeth},
Scikit-Learn \citep[version 0.21.3,][]{sklearn},
Numpy \citep[version 1.17.4,][]{numpy1, numpy2},
Matplotlib \citep[version 3.1.1,][]{matplotlib},
Mayavi \citep[version 4.7.1,][]{mayavi},
h5py (version 2.10.0, \url{https://www.h5py.org}),
Pickle \citep[version 0.7.5,][]{pickle},
Jupyter Notebook \citep[version 6.0.3,][]{jupyterNotebooks}}}

\bibliography{references}{}
\bibliographystyle{aasjournal}


\end{document}